 \documentclass[final,5p,times,twocolumn]{elsarticle}

\usepackage[numbers]{natbib}
\usepackage[hidelinks]{hyperref}
\usepackage{url}
\usepackage{etoolbox} 
\usepackage{enumitem}
\usepackage{multirow}
\usepackage{array}
\usepackage{caption}
\usepackage{graphicx}
\usepackage{array}
\usepackage{glossaries}
\usepackage{float}
\usepackage[utf8]{inputenc}
\usepackage[T1]{fontenc}
\usepackage{longtable}
\UseRawInputEncoding

\usepackage{acronym} 
\usepackage{amssymb}
\usepackage{amsmath}
\usepackage{xcolor} 
\usepackage{cleveref}
\definecolor{customcyan}{rgb}{0.0, 0.6, 0.8}
\crefformat{figure}{\textcolor{customcyan}{Fig.~#2#1#3}}
\crefformat{table}{\textcolor{customcyan}{Table~#2#1#3}}

\journal{XXX}

\begin{document}

\begin{frontmatter}

%% Title, authors and addresses

\author{Germano Rugendo Mugambi\corref{cor1}\fnref{label1,label2}}
\ead{gemuga@dtu.dk}
% \ead[url]{home page}
% \fntext[label]{}
\cortext[cor1]{Corresponding author}
\author{Nicolae Darii \fnref{label1,label3}}
\ead{nidar@dtu.dk}

\title{Methodologies for offshore wind power plants stability analysis} %% Article title

%% use optional labels to link authors explicitly to addresses:
% \author[label1,label2]{Germano Rugendo Mugambi}
% \author[label1,label3]{Nicolae Darii}
\author[label2]{Hesam Khazraj}
\author[label1]{Oscar Sabor\'{i}o-Romano}
\author[label2]{Alin George Raducu}
\author[label3]{Ranjan Sharma}
\author[label1]{Nicolaos A. Cutululis}
\affiliation[label1]{organization={Department of Wind and Energy Systems, Technical University of Denmark},
            addressline={Frederiksborgvej 399},
            city={Roskilde},
            postcode={4000},
            %state={},
            country={Denmark}}

\affiliation[label2]{organization={Vattenfall Vindkraft A/S, Denmark},
            addressline={Jupitervej 6},
            city={Kolding},
            postcode={6000},
            % state={},
            country={Denmark}}
\affiliation[label3]{organization={Siemens Gamesa A/S, Denmark},
            addressline={Elektrovej},
            city={Kongens Lyngby},
            postcode={2800},
            % state={},
            country={Denmark}}

%% Abstract
\begin{abstract}
%% Text of abstract
The development of larger \acp{OWPP} is moving towards multi-vendor setups, ultimately aiming to establish Energy hubs. These structures are characterized by installations from different vendors sharing the same connection or closely interconnected points. Control interactions among \ac{WT} converters and power systems have been detected, and this critical phenomenon can significantly impact the dynamic stability of the system. Various stability analysis methods have been proposed to analyze the interactions between \acp{OWPP} at the \ac{PoC} and the power system. However, stability studies rarely consider the complex offshore transmission system behind the \ac{PoC}. Generally, the overall \ac{OWPP} is blamed for the instability. However, since it is a complex system, it is important to understand which part of the \ac{OWPP} behind the \ac{PoC} is causing the problem or is likely to become unstable under certain conditions. Therefore, this paper provides a detailed overview of the advantages and limitations of the current system screening indexes used to design the \ac{OWPP}, and the stability analysis methods. Each method is discussed, and the appropriate methods, depending on \ac{OWPP} structure, are evaluated and discussed. The analysis indicates that a combination of time domain and frequency domain methods is necessary for enhancing the definition of stability boundaries.
\end{abstract}

%% Keywords

\end{frontmatter}

%% Use \section commands to start a section
\section*{Nomenclature}
\begin{acronym}[STATCOM]
\acro{AC}{Alternating Current}
\acro{CSCR}{Composite Short Circuit Ratio}
\acro{DC}{Direct Current}
\acro{EMT}{Electromagnetic Transients}
\acro{ESCR}{Equivalent Circuit-based Short-circuit Ratio}
\acro{FFT}{Fast Fourier Transform}
\acro{GFL}{Grid-Following}
\acro{GFM}{Grid-Forming}
\acro{GSIM}{Grid Strength Impedance Metric}
\acro{HVAC}{High-Voltage Alternating Current}
\acro{HVDC}{High-Voltage Direct Current}
\acro{HIL}{Hardware-In-the-Loop}
\acro{IBR}{Inverter-Based Resource}
\acro{IF}{Interaction Factors}
\acro{IMR}{Impedance Margin Ratio}
\acro{IP}{Intellectual Property}
\acro{PED}{Power Electronic Devices}
\acro{LTI}{Linear Time Invariant}
\acro{MIIF}{Multi-Infeed Interaction Factors}
\acro{MIMO}{Multiple-Input and Multiple-Output}
\acro{MMC}{Modular Multi-Level Converter}
\acro{MVIF}{Multi-Voltage Interaction Factors}
\acro{ODE}{Ordinary Differential Equations}
\acro{PLL}{Phase-Locked Loop}
\acro{PoC}{Point-of-Connection}
\acro{PRBS}{Pseudo-Random Binary Sequence}
\acro{RMS}{Root Mean Square}
\acro{SCR}{Short-Circuit Ratio}
\acro{SCRIF}{Short Circuit Ratio with Interaction Factors}
\acro{SCL}{Short-Circuit Level}
\acro{SISO}{Single-Input and Single-Output}
\acro{STATCOM}{Static Synchronous Compensator}
\acro{TSO}{Transmission System Operator}
\acro{VSC}{Voltage Source Converter}
\acro{OWF}{Offshore Wind Farm}
\acro{OWPP}{Offshore Wind Power Plant}
\acro{WT}{Wind Turbine}
\end{acronym}

\section{Introduction}
\label{sec:introduction}
%\subsection{General view of offshore wind}
The quest for a climate neutral by 2050 has seen tremendous yearly investments in offshore wind turbine installations worldwide. According to the Global Offshore Wind Report 2024, offshore wind installations have a 24 \% annual growth \cite{GlobalWindEnergyCouncil2024GLOBAL2024}. The International Renewable Energy Agency (IRENA) predicts nearly 500 GW of global cumulative offshore wind installed capacity by 2030 from the current 75.2 GW \cite{GlobalWindEnergyCouncil2024GLOBAL2024,PatentOffice2023OffshoreReport,RenewableEnergyAgency2023WorldPathway}. Some countries in Asia-Pacific and Europe have set ambitious targets for achieving their energy transition through offshore wind. For example, in the Esberg Declaration, Denmark, Germany, Netherlands, and Belgium have already agreed to develop a common offshore wind system in the North Sea with a capacity of 300 GW by 2050 \cite{Savaghebi2023GeneralPioneer,Author2023ModelAnalysisb}. 

With the upcoming large-scale Offshore Wind Power Plants \acp{OWPP} and energy hubs, more extensive \ac{HVDC} or \ac{HVAC} transmission is envisaged to connect multiple OWPPs at the offshore connection point. The interconnection of \ac{OWPP} clusters from different manufacturers will result in a multi-vendor offshore system. Some future planned multi-vendor offshore wind projects that will share a connection point include SOFIA and DOGGER BANK C with 100 Siemens Gamesa \& GE Haliade 14 MW wind turbines, respectively \cite{RTEInteractionFarms}. 

%\subsection{Challenges of present and future offshore interconnections}
Typically, an \ac{AC} offshore system is characterized by low inertia, low damping, and possible interaction between an \ac{HVDC} converter and wind turbine converters for an \ac{HVDC}-connected \ac{OWPP} \cite{AbdalrahmanAdil2016DolWin1Farms}. These challenges are exacerbated if the control systems of the multiple converters are not identical \cite{Lu2017LettersFilter}. The interactions could be in the form of harmonics, oscillation due to voltage instability, or system trip due to overcurrents. The nature of harmonics depends on the amount of power generated by wind turbines and the topology of the offshore \ac{AC} network \cite{AbdalrahmanAdil2016DolWin1Farms}. Hence, for the envisaged interconnections where large \acp{OWPP} from different manufacturers share connection points offshore, it is important to investigate possible adverse control interactions among hundreds of wind turbine converters whose control characteristics are different and their impact on the offshore \ac{AC} grid.

Another recent development that may introduce stability challenges is the collector system voltage, which is planned to increase from 66 kV to 132 kV. The voltage increase may make the use of an offshore transformer redundant in some cases. In such scenarios, several very large offshore wind power plants from different manufacturers will be connected to the same hub without a dedicated offshore plant transformer, raising concerns about the dynamic stability of the offshore grid due to possible interactions between hundreds of converter-interfaced wind turbines. TenneT is already leading on this front, as the upcoming Ijmuiden Ver wind farm and DolWin5 grid connection will have no wind farm transformer. The wind power plants will be connected directly to the 66-kilovolt switchyard at the offshore converter station \cite{TenneT2023AnnexesAgreement}.

Recent real events, such as the increase of potential \acp{OWPP} disconnections and consequent blackouts in UK \cite{WhatGuardian} or low-frequency oscillatory effects induced by \ac{IBR} interactions \cite{Rudnik2022AnalysisSources} have raised the interest of industry and \acp{TSO} in the application of different stability analysis methods to identify and de-risk the \acp{IBR} penetration and operation. Several methods are investigated in literature and, more specifically, applied to \ac{OWPP} case studies. These methods are classified into Eigenvalue-based, frequency-domain, and time-domain simulations. The \acp{TSO} have widely used the impedance-based method because of its well-defined stability metrics and provision of efficient ways from a system perspective. However, measurements must be made in time domain simulations to obtain converter impedance for a wide range of frequencies. 

Converter manufacturers have used the eigenvalue-based stability analysis method to investigate converter-related stability \cite{BAKHSHIZADEH2019GridPlants}. In some instances, a combination of these methods is required to analyze some interaction phenomena fully. For example, a combination of impedance-based method and time domain simulation was used to analyze an interaction between wind power plant converters and offshore \ac{HVDC} converter \cite{C.Buchhagen2016HarmonicTSO}. The majority of the studies have emphasized investigating \ac{OWPP} interactions with the onshore system and their impact on the power system stability. The stability analysis of the offshore system, i.e. the impact of wind turbine converter control interactions on the offshore system stability, has not been addressed. 

A recently published CIGRE technical brochure evaluates these methods using a proposed benchmark system, including a comparison of their industry maturity. Further, the brochure provides procedures and guidelines for industry and academia on how to perform small-signal stability studies in modern converter-dominated power systems \cite{WGC4.492024Multi-frequencyBROCHURE}. However, as mentioned above, the analysis is focused on the impact of overall power system stability by power converters. Additionally, not all the methods can be applicable in the case of multi-vendor \ac{OWPP} clusters. 

Stability issues due to control interaction between converters of different manufacturers have been observed in the first European multi-vendor parallel connected \ac{HVDC} converters in grid-forming operation, the Johan Sverdrup project. The project involves a \ac{MMC} \ac{HVDC} link and a 2-level \ac{VSC} \ac{HVDC} link from different vendors operating in parallel to supply an offshore oil field \cite{RTDSTechnologies2022ENSURINGTESTING}. The interactions were investigated using detailed vendors, black-boxed replicas of their control and protection hardware, and \ac{HIL} real-time simulations. The method is unsuitable for offshore grids with more converters because of the many possible configurations to analyze \cite{VordemBergeSOLUTIONSSYSTEMS}.

The main objectives of this paper are:
\begin{enumerate}[label=\roman*.] % i.
\item To provide a detailed analysis of the methodologies used to study control interactions in converter-dominated power systems and critically evaluate the most appropriate methods for studying interactions in the case of large offshore interconnected multi-vendor \acp{OWPP}.
\item To evaluate the advantages and limitations of each method, providing insights into their applicability for small signal stability studies in complex multi-vendor \acp{OWPP} systems. 
\item To present an overview of the recent development in stability challenges associated with large offshore wind clusters to provide readers with thorough insights into the stability analysis process.
\item To provide future research outlook of multi-vendor offshore wind.
\end{enumerate}

The rest of the paper is organized as follows: Section \ref{sec:topologies} describes the current and future offshore wind power plant's topologies. The detailed analysis of stability analysis methods, including their advantages, limitations, and applicability, is presented in Section \ref{sec:methods}. In Section \ref{sec:discussion}, a summary of the author's perspective on the methods and future research prospects are highlighted. Finally, conclusions are provided in Section \ref{sec:conclusion}.

\section{Technologies of Offshore Wind Power Plants}
\label{sec:topologies}

The controllability, flexibility, and redundancy of offshore transmission systems depend on the topology and technology utilized. There are two significant technologies in grid-connected \acp{OWPP} i.e., \ac{HVAC} and \ac{HVDC}. Due to extensive knowledge and continuous technological evolution, \ac{HVAC} systems have been the dominant technology in power systems. However, depending on the project, \ac{HVAC} could face limitations in meeting substantial power capacity and long-distance transmission requirements mainly due to the high cable reactive-capacitive losses. In order to overcome this limit, inductive reactors placed in parallel to the \ac{HVAC} cable ends are generally the first adopted solution  \cite{Elliott2016ABritain}. 

Generally, \ac{HVAC} can be easily employed for short distances, i.e. towards the higher end of 10s of km, after which it will require compensation by shunt reactors. \cite{Dakic2021HVACTransmission,Gomis-Bellmunt2011Voltage-currentFarms,Nazir2022Multi-terminalCoast}. However, the inductive compensator is also associated with higher costs for the platforms, \ac{HVAC} cable's interruption, limits on the maximum inductor's size thus the need for multiple of them and high inrush currents and potential converter's saturation \cite{Liao2023StudyGrid, Wiechowski2008SelectedTSO}.

% \end{enumerate}
In addition, it is possible to use Static VAR, Synchronous Condensers, or \acp{STATCOM} placed at the onshore end to control the voltage (thus the reactive power generated from the \ac{HVAC} cable) \cite{ Wu2024GridTransmissionb, Nguyen2021ACondensers}.

When these solutions are no longer technically or economically feasible, opting for the point-to-point \ac{HVDC} topology is possible. 
The \ac{HVDC} can ameliorate some challenges connected with \ac{HVAC} topology in evacuating large amounts of power from large offshore wind power plants in deep waters. Therefore, depending on the project requirements, it is possible to shift to the other technology if necessary \cite{Rodrigues2015TrendsProjects,Ansari2020MMCSchemes,Korompili2024ReviewIntegration}. The HVAC and HVDC offshore transmission systems can be realized in different topologies, as described below.

\subsection{Current Topologies}
\begin{figure}[h!]
    \centering
    \includegraphics[width=1\linewidth]{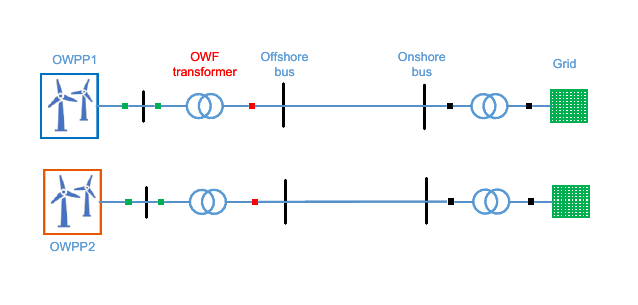}
     \caption{\ac{HVAC}-connected \ac{OWPP}}
    \label{fig:HVAC_ptp}
\end{figure}

\begin{figure}[h!]
    \centering
    \includegraphics[width=1\linewidth]{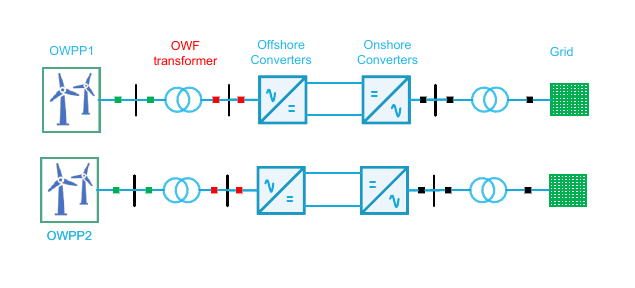}
    \caption{Point-to-point connection of \ac{HVDC}-connected \ac{OWPP}}
    \label{fig:HVDC_ptp}
\end{figure}

Before Nan'Ao  3-terminal and Zhoushan's 5-terminal projects in China, all offshore wind projects were built as point-to-point connections \cite{Bathurst2015DeliverySystem,NRElectricCo.Ltd2015WorldsLinks}. The bespoke nature of offshore projects leads to each project having a dedicated offshore platform and transmission infrastructure to the onshore substation as illustrated in Figs. \ref{fig:HVAC_ptp} and \ref{fig:HVDC_ptp}. Having each project have its infrastructure has led to delays in project completion due to resistance from the general public and long approval times due to limited areas on the shore where onshore infrastructure has to be put on, jurisdictional issues, and stakeholder interests. At the scale of large \ac{OWPP} projects and anticipated energy hubs, delays will result in increased project costs. Additionally, having several cables laid on the seabed could lead to loss of biodiversity and negative environmental impacts \cite{Savaghebi2023GeneralPioneer,CutululisGeneralSources}.

Failure or maintenance of a transmission line or converter station could lead to a complete loss of connection, potentially causing power supply interruptions. The topology is also limited to flexibility in expansion. Integrating additional power plants will require significant modifications or additional infrastructure, leading to higher costs and complexity. 
Therefore, when the project necessities for \ac{OWPP} with higher wind power productions, it needs a consequent reliability requirement from the interconnection safety point of view and newer topologies such as the \ac{HVDC} Multi-terminal could provide the required conditions according to the literature.


\subsection{Potential future topologies}
Although the construction of Nan'Ao and Zhoushan multi-terminal projects marked significant advancements in \ac{HVDC} technology; their development circumstances differ significantly from the competitive environment currently observed in Europe and other Western countries. In Europe, converter manufacturers typically provide high-voltage equipment and the Control and Protection (C\&P) system that governs its operation. In contrast, the Chinese projects adopted a different approach, with multiple vendors supplying the \ac{HVDC} valves and their respective valve-based controls, while a single supplier managed the overall C\&P system for all converters involved. This regulatory structure diverges from European practices, where the value and \ac{IP} associated with \ac{HVDC} converters predominantly reside in the control algorithms owned by the suppliers \cite{DCSolutions}.

The core \ac{IP} of power converter manufacturers is the converter control algorithm. To promote competitiveness and innovation, the industry is expected to remain like this, especially in Europe. This implies that interconnected \ac{OWPP} clusters will typically be multi-vendor multi-converter systems. The challenge paused with the interconnection is the interoperability of the converters which threaten the dynamic stability of the offshore \ac{AC} grid which is characterized by weaker sources. The converter's dynamic behavior is highly dependent on topology and control design. So far, there are no multi-vendor \ac{OWPP} clusters in operation, but the TenneT Ijmuiden Ver wind farm is being developed with the possibility of the three wind farms being from different manufacturers sharing an offshore converter platform \cite{TenneT2023AnnexesAgreement}.

Different topologies can be realized depending on whether the \acp{OWPP} are interconnected on the \ac{AC} or \ac{DC} sides. The choice of interconnection type relies heavily on various factors, including capital investment, potential for future expansion, availability of technology, and other pertinent considerations. The different topologies of interconnected \acp{OWPP} are described below.

\subsubsection{Multi-infeed topology }
\begin{figure}
    \centering
    \includegraphics[width=1\linewidth]{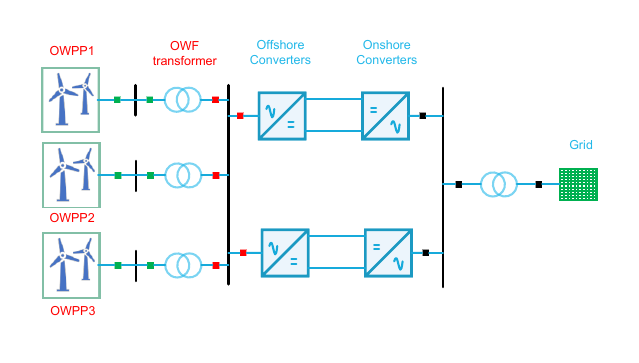}
    \caption{Multi-infeed offshore \ac{HVDC} topology}
    \label{fig:multi-infeed}
\end{figure}

This topology interconnects multiple \acp{OWPP} on the \ac{AC} side, allowing for shared utilization of the converter platform. The \ac{HVDC} converters are directly linked to the onshore converters. Notably, the same vendor provides all the control and protection involved in this configuration to ensure uniformity in their control characteristics. Despite the C\&P being from the same vendor, there could be the challenge of technological interoperability. For example, operational compatibility between 2-level \acp{VSC} and \acp{MMC} from the same manufacturer \cite{ENTSO-E2021WorkstreamDevices}. In addition, in Europe, the common practice is to have different manufacturers for the offshore-\ac{MMC} and \acp{WT} in the same \ac{OWPP}. The topology is illustrated in Fig. \ref{fig:multi-infeed}. When a single \ac{HVDC} line is used, the topology is considered the most economical solution for the standard 2 GW \ac{OWPP} clusters.

\subsubsection{Multi-infeed Multi-vendor topology}
\begin{figure}
    \centering
    \includegraphics[width=1\linewidth]{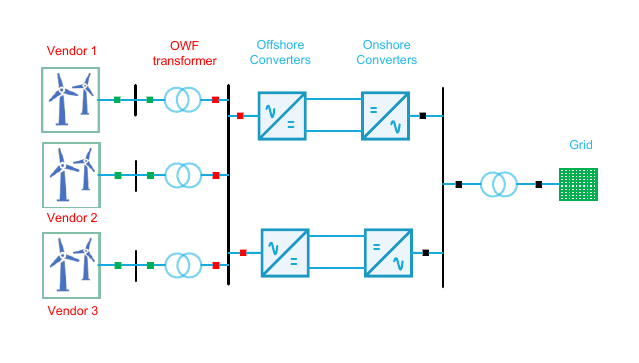}
    \caption{Multi-infeed Multi-vendor \ac{HVDC} topology}
    \label{fig:MIMV}
\end{figure}

This topology is similar to multi-infeed, except the \ac{OWPP} clusters C\&P are from different vendors. For example, in Fig. \ref{fig:MIMV}, the power plants using converters from three different manufacturers are sharing an offshore connection point. Interconnection of clusters with different controller characteristics can lead to manufacturer interoperability issues due to possible negative interactions between converters \cite{ENTSO-E2021EuropeanInteroperability}. This implies that the technology could be the same but from different manufacturers.

\subsubsection{Multi-terminal topology}
\begin{figure}
    \centering
    \includegraphics[width=0.8\linewidth]{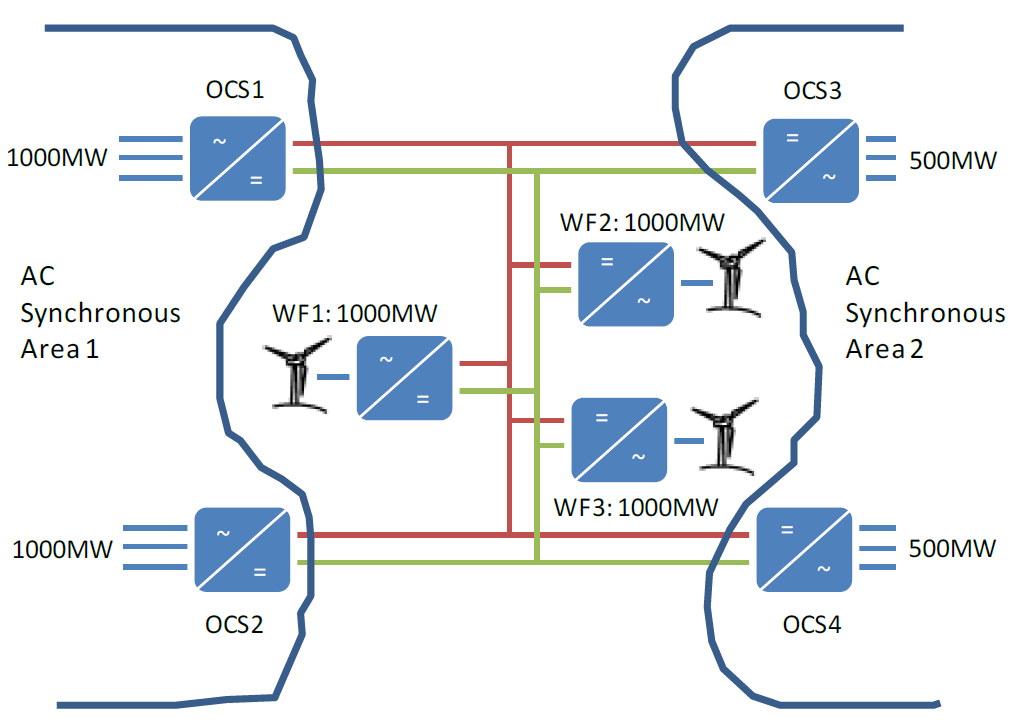}
    \caption{Multi-terminal Multi-vendor \ac{HVDC} (\ac{DC} grid) \cite{DCSolutions}}
    \label{fig:MTMV}
\end{figure}

The interconnection of \ac{OWPP} clusters on the \ac{DC} side leads to multi-terminal \ac{HVDC}, as shown in Fig. \ref{fig:MTMV}. The control and protection systems for the clusters could be from the same vendor or different vendors. If they are the latter, the topology is Multi-vendor Multi-terminal (MTMV) \ac{HVDC}. The development of multi-terminal \ac{HVDC}/\ac{DC} grids is expected to accelerate the growth of offshore wind and minimize the problems of point-to-point \ac{HVDC} systems. They are expected to offer more redundancy and increased reliability than their point-to-point counterpart. If one terminal fails, the system can continue to operate using other paths, thus minimizing downtime and ensuring a stable power supply. Furthermore, multi-terminal \ac{HVDC} systems are flexible to expansion, allowing integration of future \acp{OWPP} or grid infrastructure without the need for additional infrastructure investment. This flexibility of expansion also minimizes environmental impacts. 

In addition, interconnecting the same \ac{OWPP} on different synchronous areas with Hybrid AC/DC connections may be possible. The next connection requirements for \ac{HVDC} take into account the possibility of having both \ac{HVAC} and \ac{HVDC} branches departing from the same \ac{OWPP} allowing the synchronization of the \acp{WT} with a specific synchronous area. Concurrently, it sends the power through the \ac{HVDC} link towards a different synchronous area \cite{GridAmendment}. This topology may increase the reliability of the \ac{OWPP} since it may facilitate the black-start of the plant from the \ac{HVAC} connected area, for example, by energizing the offshore components and launching the Offshore-\ac{MMC} converters' cooling systems. 

\subsubsection{Challenges of multi-terminal \ac{HVDC} systems}
While multi-terminal \ac{HVDC} offers numerous advantages, it is also faced with complexity. Designing the coordination of multiple terminals and converters is complex. In multi-terminal multi-vendor (MTMV) systems, the issue of interoperability among \acp{PED} from different manufacturers whose control characteristics are different is a hot topic in industry and academia. At least in Europe, offshore wind is seen as one of the cornerstones of achieving net zero emissions, meaning that components with \ac{PED}, such as wind turbine converters, \ac{HVDC} converters, etc., must be interoperable to achieve the much-needed installations.

Earlier this year, the InterOPERA project was started to facilitate the operation of MTMV \ac{HVDC} grids in Europe. The project mainly aims to develop technical standards and frameworks for \ac{HVDC} projects \cite{EuropeanClimate2023SuccessfulProject}. Recent developments in offshore collector system voltage that might make offshore transformers at the wind power plant level obsolete present more interoperability challenges. The core of VSC or MMC converters are their controls. The interactions among inverters (controls) and power systems depend on the coordination of the controller tuning or the sensitivity to the power systems parameter and their change in time. Therefore, there is a great need to develop control methods to ensure dynamic stability and tools to analyze possible interactions in \ac{HVDC} systems.

\section{Stability Analysis Methods}
\label{sec:methods}
 Different methods have been used in the literature to analyze the possible interaction of \acp{IBR} with the \ac{AC} network they are connected to and other power electronic devices in close proximity. This section presents a detailed analysis of the methods, including their limitation and practicability for multi-vendor \acp{OWPP}. These methods are summarized in Tables \ref{tab:comparison_table1} and \ref{tab:comparison_table2}.
\subsection{System Screening}
Prior to the connection of new assets such as \acp{OWPP} to the power system, an initial screening is done, which does not involve detailed analytical and numerical analysis. The purpose of the screening is to give an overview of potential interactions with other generating units or system components. The interactions are mainly influenced by short circuit power and the electrical distance between system equipment \cite{Conseilinternationaldesgrandsreseauxelectriques.ComitedetudesB4.2008SystemsInfeed}. However, to quantify the problematic scenario or instability captured during the screening process, a detailed analytical or numerical analysis is needed \cite{WGC4.492024Multi-frequencyBROCHURE}. The small-signal screening techniques employed by industry and academia are discussed below.
\subsubsection{Interaction Factors (IF)}
The interaction factor is a metric that provides the voltage sensitivity of a particular bus due to a disturbance at another bus. It is given as a ratio between a disturbance at one bus and the reaction at another bus at some electrical distance from the first bus. Interaction factors are classified into \ac{MIIF} and \ac{MVIF} depending on the disturbance used. \ac{MIIF} is obtained when reactive power is injected to cause a voltage disturbance, while an active power injection is used for \ac{MVIF}. \ac{MIIF} and \ac{MVIF} are equivalent as indicated by equation \eqref{eqn:MIIF}, however, \ac{MIIF} is not applicable for devices operating under constant \ac{AC} voltage control \cite{Conseilinternationaldesgrandsreseauxelectriques.ComitedetudesB4.2008SystemsInfeed}.

\begin{equation}
    \text{MIIF}_{2,1} \equiv \text{MVIF} = \frac{\Delta v_2}{\Delta v_1}
    \label{eqn:MIIF}
\end{equation}

Where $\Delta v_1$ and $\Delta v_2$ are the voltage perturbations at bus 1 and 2 respectively. A voltage disturbance can be created by switching a fictitious inductive shunt element at bus 1. The \ac{MIIF}/\ac{MVIF} values range from 0 to 1, where 0 indicates an infinite distance between buses 1 and 2 while 1 indicates the two buses are the same bus \cite{Conseilinternationaldesgrandsreseauxelectriques.ComitedetudesB4.2008SystemsInfeed}. High values of \ac{MIIF} indicate a potential risk of instability/interactions and a need for further detailed investigation. \ac{IF} identify potential issues by analyzing passive network components between power electronic devices, and they can provide only limited information on the controller structure. For systems with several converters in close proximity, multiple \acp{MIIF} are calculated, forming a matrix. Typically, the \ac{MIIF} matrix is asymmetric because the values between two converter buses depend not only on the intervening impedance but also on the shunt impedance at each converter bus. Additionally, the diagonal elements are unity. An example of the \ac{MIIF} matrix for three converters is shown in Table \ref{tab:miif_table}.

\begin{table} [h!]
    \caption{An example of \ac{MIIF} Matrix for three converters}
    \centering
    \resizebox{0.48\textwidth}{!}{
    \begin{tabular}{c c c c c }
        \hline
        \multirow{2}{*}{\textbf{MIIF Table}} & \multirow{2}{*}{} & \multicolumn{3}{c}{\textbf{Relative converter bus \ac{AC} voltage change}} \\ \cline{3-5}
         & & \textbf{Bus 1} & \textbf{Bus 2} & \textbf{Bus 3} \\ \hline
        \multirow{3}{*}{\begin{tabular}[c]{@{}c@{}}\textbf{Voltage}\\ \textbf{disturbance}\\ \textbf{injection bus}\end{tabular}} 
        & \textbf{Bus 1} & \(\text{MIIF}_{1,1} = \frac{\Delta v_1}{\Delta v_1}\) & \(\text{MIIF}_{2,1} = \frac{\Delta v_2}{\Delta v_1}\) & \(\text{MIIF}_{3,1} = \frac{\Delta v_3}{\Delta v_1}\) \\ \cline{2-5} 
         & \textbf{Bus 2} & \(\text{MIIF}_{1,2} = \frac{\Delta v_1}{\Delta v_2}\) & \(\text{MIIF}_{2,2} = \frac{\Delta v_2}{\Delta v_2}\) & \(\text{MIIF}_{3,2} = \frac{\Delta v_3}{\Delta v_2}\) \\ \cline{2-5} 
         & \textbf{Bus 3} & \(\text{MIIF}_{1,3} = \frac{\Delta v_1}{\Delta v_3}\) & \(\text{MIIF}_{2,3} = \frac{\Delta v_2}{\Delta v_3}\) & \(\text{MIIF}_{3,3} = \frac{\Delta v_3}{\Delta v_3}\) \\ \hline
    \end{tabular}
    }
    \label{tab:miif_table}
\end{table}

It is recommended that all converter systems whose \ac{MIIF} is above 0.15 should be investigated further. Table \ref{tab:miif_table} above shows that the \ac{MIIF} values describe the coupling between two converter buses without considering their ratings. Reference \cite{Conseilinternationaldesgrandsreseauxelectriques.ComitedetudesB4.2008SystemsInfeed} recommends taking into account the size of the converter, as a large converter will have more influence on the voltage sensitivity at a particular bus than a smaller one. Hence, a relative weighted \ac{MIIF} is desired, which is a product of \ac{MIIF}, and the rated \ac{DC} power of the new link, calculated as a percentage of the rated \ac{DC} power of the existing link. Relative weight \ac{MIIF} of less than $15\%$ indicates low chances of interaction, while values between $15\%$ and $40\%$ indicate high potential of converter interaction. The product values above $40\%$ indicate a high potential for interactions.

\ac{MIIF} and \ac{MVIF} being scalar values, they provide a quick overview of the system, but they cannot identify specific interactions. Hence, it may result in the mischaracterization of the system. \ac{IF} are calculated within the fundamental frequency; thus, control interactions across the controller bandwidth are not characterized \cite{Henderson2024GridSystems}. Furthermore, since the interaction factor is a scalar value, further small-signal analysis, such as eigenvalue analysis, cannot be performed using the information from \ac{IF}. Moreover, it can be interpreted that \ac{IF} measures the effect of the network impedance, which could be obtained through a short-circuit ratio. Therefore, it is essential to use a screening technique whose output enables more in-depth analysis in further studies. 

\subsubsection{System strength}
The traditional power system strength is defined based on the available short-circuit current, vulnerability to voltage perturbations, and maximum power transfer \cite{Henderson2024GridSystems,AEMC2017,system_ISGT}. To capture the above three aspects, metrics such as \ac{SCL} and \ac{SCR} have been used. The metrics \ac{SCR} and \ac{SCL} refer to the volume of extra fault current compared to nominal current when a three-phase fault occurs at a specific point in the system. In a conventional power system with many synchronous generators, \ac{SCL} is expressed as \ac{SCR} which translates to system strength. The \ac{SCR} relates to the passive impedance between the specific point on the network and the source of the fault current, hence not sufficient to characterize a system dominated by converters \cite{Henderson2022AnalysisSystems,Zhu2024}. To ameliorate the \ac{SCR} inadequacies for converter-dominated systems, two new metrics called Grid strength impedance metric \cite{Henderson2024GridSystems} and Impedance Margin Ratio \cite{Zhu2024} have been proposed. 

\paragraph{Short circuit ratio (SCR)} \mbox{}\\
 Since \ac{SCR} is determined by physical line impedances, it describes the electrical distance between a point on the network and a stiff voltage source. A higher value of \ac{SCR} is characterized by low impedance and shorter electrical distance, resulting in a strong system. A stiffer system provides good damping to the voltage disturbances. On the other hand, long electrical distances result in large impedances and low \ac{SCR}. A system with a low \ac{SCR} value is termed a weak system, with poor voltage damping, i.e., rapid changes in system voltage with changes in injected or absorbed reactive power. Classification of system strength in terms of SCR is summarized in Table \ref{tab:SCR_table} \cite{Ghimire2023Small-SignalCondensers}.

\begin{table}[h!]
\caption{Classification of system strength in terms of SCR}
\centering
\begin{tabular}{p{3.5cm} p{3.5cm}} 
\hline
\textbf{Grid Case} & \textbf{SCR value} \\ \hline
Very weak Grid & $<$ 2 \\ 
Weak Grid & $2 < \text{SCR} \leq 3$ \\ 
Strong Grid & $>$ 3 \\ \hline
\end{tabular}
\label{tab:SCR_table}
\end{table}

 In the converter-dominated system of today, \ac{SCR} no longer provides an accurate measure of the electrical distances between points on the network and stiff voltage sources. This is because the fault current contribution from converters is not governed by the passive impedance seen between the converter terminal and the network point of study. While the \ac{SCR} is still useful for assessing fault conditions, it presents an incorrect characterization of the system stiffness during normal operation. In addition, \ac{SCR} is an indicator of the p.u. impedance of the system at the fundamental frequency; hence, it does not consider resonances that may be present in the grid and interactions in case multiple \acp{IBR} in a multi-infeed system \cite{WGC4.492024Multi-frequencyBROCHURE,Henderson2024GridSystems, Zhu2024,Henderson2022ExploringConverters}. To avoid the wrong definition of \ac{SCR} and system strength for a system dominated by \acp{IBR},  the suggestion is to ignore the contribution of \ac{IBR} or use an alternative definition \cite{Henderson2024GridSystems,Zhu2024}.

 Alternative \ac{SCR} definitions for systems with high penetration of converters include:
 
\begin{enumerate}[label=\Roman*.]
    \item \textit{Equivalent circuit-based short-circuit ratio (ESCR)}

The difference between ESCR metric and conventional \ac{SCR} is that it considers all the physical impedances on the network at the point of interconnection as illustrated in equation \eqref{eqn:ESCR}. However, it does not consider the converter control impedances \cite{Conseilinternationaldesgrandsreseauxelectriques.ComitedetudesB4.2016ConnectionNetworks}.
    \begin{equation}
        \text{ESCR} = \frac{1}{Z_{sys, PU}}
        \label{eqn:ESCR}
    \end{equation}

Where $Z_{sys, PU}$ is the network pu impedance at the point of interconnection.

    \item \textit{Composite short circuit ratio (CSCR)}
    
The CSCR calculates an aggregate \ac{SCR} for multiple \acp{IBR} instead of each resource, as in the traditional \ac{SCR} method. All \acp{IBR} of interest are assumed to be connected to a single bus and the composite short-circuit MVA at the common bus is calculated but without fault current contribution from \acp{IBR}. The \ac{CSCR} is calculated by equation \eqref{eqn:CSCR}.
    \begin{equation}
        \text{CSCR} = \frac{\text{CSC}_{MVA}}{\text{MW}_{VER}}
        \label{eqn:CSCR}
    \end{equation}

Where $CSC_{MVA}$ is the fault level contribution without \acp{IBR} and $MW_{VER}$ is the nominal power rating of the connected \acp{IBR} \cite{Henderson2024GridSystems,Conseilinternationaldesgrandsreseauxelectriques.ComitedetudesB4.2016ConnectionNetworks,NERC2017NERCSystems}. Although the CSCR approach give a more accurate estimate of the system strength compared to the \ac{SCR} metric, it assumes there is a strong electrical coupling between the \acp{IBR}.

    \item \textit{Weighted short circuit ratio (WSCR)}
    
The weighted short circuit ratio approach is similar to \ac{CSCR}, but it is possible to analyze key points on the grid using multiple buses as described in equation \eqref{enq:WSCR}. 

    \begin{equation}
        \text{WSCR} = \frac{\sum_i^N{\text{SCMVA}_i\cdot P_{RMW_i}}}{(\sum_i^N{P_{RMW_i}})^2}
        \label{enq:WSCR}
    \end{equation}

Where ${SCMVA}_i$ is the short circuit capacity at bus i without current contribution from \acp{IBR} and $P_{{RMW}_i}$ is the rated output power of the \ac{IBR} to be connected to bus i \cite{NERC2017NERCSystems,Evaluate2014}.

    \item \textit{Short circuit ratio with interaction factors (SCRIF)}

The \ac{SCR} definitions described so far account for impedances between the considered \acp{IBR}. Although the approach considers each \ac{IBR} connected to a certain bus, some aspects, such as voltage stiffness introduced by \ac{IBR} control schemes such as \ac{GFM}, are not presented. To capture the voltage sensitivity, voltage deviations must be integrated into the \ac{SCR} computations. Hence, \ac{SCRIF} captures these voltage deviations using interaction factors. \ac{SCRIF} is determined as shown in equation \eqref{eqn:SCRIF}.

    \begin{equation}
        \text{SCRIF}_i = \frac{S_i}{P_i+\sum_j{IF_{ij}\cdot P_j}}
        \label{eqn:SCRIF}
    \end{equation}
    
Where $S_i$ and $P_i$ are the short circuit power and nominal power rating at bus i and $P_j$ is the nominal power rating at bus j. ${IF}_{ij}$ is the change in bus voltage at bus i (${\mathrm{\Delta V}}_i$) for a change in voltage at bus j ($\mathrm{\Delta}V_j$) and is computed by equation \eqref{eqn:IF}.

    \begin{equation}
        IF_{ij} = \frac{\Delta V_i}{\Delta V_j}
        \label{eqn:IF}
    \end{equation}

A stiff voltage source at bus i results in a lower value of \ac{IF} and, consequently, high \ac{SCRIF}. This means that bus i will be able to damp voltage perturbations better and hence more stable. \ac{SCRIF} has the advantage of being adapted for any configuration of multiple \ac{IBR}. However, it is still faced with the drawback that it disregards the converter control impedance, as it only represents the power ratings and line impedances. Since the converter control components are not captured, \ac{SCRIF} will underestimate the system strength when grid-forming \acp{IBR} are included \cite{Henderson2024GridSystems, Henderson2022ExploringConverters, NERC2017NERCSystems}. 
\end{enumerate}

\paragraph{Grid strength impedance metric (GSIM) and Impedance Margin Ratio (IMR)} \mbox{}\\
Notably, the conventional methods used to determine the system's strength are unable to differentiate between control schemes, such as \ac{GFL} and \ac{GFM}, which have different contributions to voltage stiffness in a bus \cite{Henderson2024GridSystems,Zhu2024}. 
Coupling voltage stiffness and fault current into one metric is not valid for a converter-dominated system. For example, during quasi-steady state conditions, voltage stiffness from \ac{IBR} can increase without increased fault current. Therefore, to correctly characterize the system that is dominated by \ac{IBR}, two new metrics called \ac{GSIM} and \ac{IMR} have been proposed \cite{Henderson2024GridSystems,Zhu2024}. \ac{GSIM} calculates the system strength independently from the short circuit level. The metric obtains equivalent converter output impedance at a quasi-steady state. The equivalent impedance comprises the physical impedance as well as the control architecture. The impedance can be obtained using frequency sweeps/scans at the frequency of interest, which makes \ac{GSIM} applicable for black box models.

Once the base and system impedances are obtained, the \ac{GSIM} is obtained as shown in equations \eqref{eqn:GSIM1} and \eqref{eqn:GSIM2}.

\begin{equation}
\begin{bmatrix}
\text{GSIM}_q(s) \\
\text{GSIM}_d(s) 
\end{bmatrix} = \lambda ( Y_{sys}(s) \cdot \lambda(Z_b(s))
\label{eqn:GSIM1}
\end{equation}

\begin{equation}
    \text{GSIM}(s) = \sqrt{\frac{\textbf{GSIM}_d(s)^2 + \text{GSIM}_q(s)^2}{2}}
    \label{eqn:GSIM2}
\end{equation}

Where $\lambda(Y_{\mathrm{sys}}(s)$ and $\lambda\left(Z_b\left(s\right)\right)$ are system admittance and base impedance eigenloci, respectively, while ${GSIM}_d(s)$ and ${GSIM}_q(s)$ are the d and q-axis \ac{GSIM} components. 

The \ac{IMR} metric is defined as the ratio between the maximum permissible change in the \ac{IBR} impedance at the \ac{PoC} and the initial impedance value. This allowable change is established to prevent an oscillatory mode from shifting into the right-half plane (RHP). The metric is calculated from the whole-system admittance model based on the damping factor of the eigenvalues at \ac{PoC}. Significant variations of eigenvalues due to small changes in impedance indicate a high likelihood of the mode moving to the RHP.

In the context of offshore \ac{AC} grids, which are almost $100\%$ converter-dominated, traditional methods used to determine the system strength are not valid. Hence, \ac{GSIM} and \ac{IMR} provide a possible solution, as the whole converter behavior across a frequency range will be important in characterizing such a system. Determining the system's strength correctly is important for its stability and control. For example, grid-following converters in weak grids lead to stability issues because of voltage deviations caused by large network impedance \cite{Davari2017RobustDynamics}. On the other hand, grid-forming converters enhance system strength, meaning they can offer solutions for weak grids \cite{Henderson2022ExploringConverters}. That is why using a system metric that correctly captures the converter behavior for a converter-dominated network is important.

\subsection{Frequency domain analysis methods}

Frequency domain analysis is a technique derived from classic control theory, where time-domain signals are studied in the frequency domain. The reason for this representation is the possibility of simplifying the analysis by transforming linear \acp{ODE} into algebraic equations via Laplace transform. The frequency domain offers a global view of the system's behavior and stability. Generally, with linearized models, it is possible to move to the frequency domain via analytical representation. However, this is not always possible when the system is too complex or black-box (like \ac{IP}-protected devices). 

\subsubsection{Transfer-function based}
The Transfer-function method represents linear systems in fractions as shown in equation \eqref{eqn:tfs}, where the polynomial at the denominator encloses the key information about the stability and performance of the system. Its accuracy is proportional to the value of N.
\begin{equation}
    G(s) = \prod^N_{i=1}{\frac{1}{s-(\sigma_P + j\omega_P)}}
    \label{eqn:tfs}
\end{equation}
Where $s$ is a complex parameter, $\sigma_p$is the real part of the pole and $\omega_p$ is its natural frequency. 

The method is becoming popular among \ac{IBR}, specifically for \ac{OWPP}, since the core of the converters is its control loops, which can be represented easily as transfer functions \cite{Harnefors2007ModelingMatrices}. The system is divided into smaller transfer functions and connected to obtain the full system representation \cite{Zhang2019StabilityModeling}, and formulated in $dq$ or $\alpha\beta$ frames \cite{Bakhshizadeh2016CouplingsConverters}.  In addition, the method gives more information about the phenomena than Phasor studies, which focus on targeted frequencies; instead transfer function covers a broad range of frequencies. It is also possible a, more complex, multivariable formulation.

From the transfer function denominator's polynomial, it is possible to define if the system is stable based on the poles' sign and tune the system's performance via pole placement techniques. For example, M. Zhao \textit{et al.} \cite{Zhao2016VoltageGrid} have used the transfer function representation to represent a \ac{VSC} converter for a \ac{WT} by modeling voltage and current controls, more explicitly depending on filters and the grid strength. L. Fan \cite{Fan2019ModelingGrids} modeled a Type-4 \ac{WT} via transfer function, explicating \ac{PLL} influence, power level, and voltage controller on the stability and low-frequency oscillations.

Grey box transfer function, for example, could synthesize a scan of an \ac{OWPP} \ac{HVDC} \cite{Amin2019AImpedance}, and it is possible to derive the transfer function $G(s)$ directly from a system in state-space notation as illustrated in equation \eqref{eqn:tfs2}, for example, from state-space models of \ac{VSC}-\ac{HVDC} \cite{Prieto-Araujo2011MethodologyFarms}. 
\begin{equation}
    G(s) = \textbf{C}(s\textbf{I}-\textbf{A})^{-1}\textbf{B}+\textbf{D}
    \label{eqn:tfs2}
\end{equation}
Where $\textbf{A}$, $\textbf{B}$, $\textbf{C}$, $\textbf{D}$ are the matrices that composes the state space representation and $\textbf{I}$ is the identity matrix. 

When analytical models are absent, Vector Fitting is an option \cite{Li2022ImpedanceApplication}, where a predefined factorized transfer function is tuned to copy the continuous frequency response of a system. The more poles are considered, the more accurate the representation will be. For example, typical cable models in the frequency domain consist of a single $\pi$-section conduction; however, in \cite{DArco2019Time-Invariant-sections}, it was proved that multiple $\pi$-section in the frequency domain could correctly characterize the cable's dynamic. When numerous poles are used, it results in complex interpretation and tuning. The method is no longer valid when the system becomes non-linear, for example, during the converter controllers' saturation. Furthermore, if the transfer function is vector fitted from black-box models, the starting parameters should be decided holistically based on the experience \cite{Li2022ImpedanceApplication, Gong2018ParametricConverters}.

\subsubsection{Impedance-based}
An impedance-based method was first used in 1976 by J.M. Undrill and T.E. Kostyniak to analyze sub-synchronous oscillations of a network using generator and transmission network impedances \cite{J.M.UndrillT.E.Kostyniak1976SubsynchronousAnalysis}. R.D. Middlebrook also used the approach in the same year to design \ac{DC}-\ac{DC} converter input filters \cite{Liao2020Impedance-BasedPoles}.

\begin{figure*}
    \centering
    \includegraphics[trim=10pt 10pt 10pt 10pt, clip, width=1\linewidth]{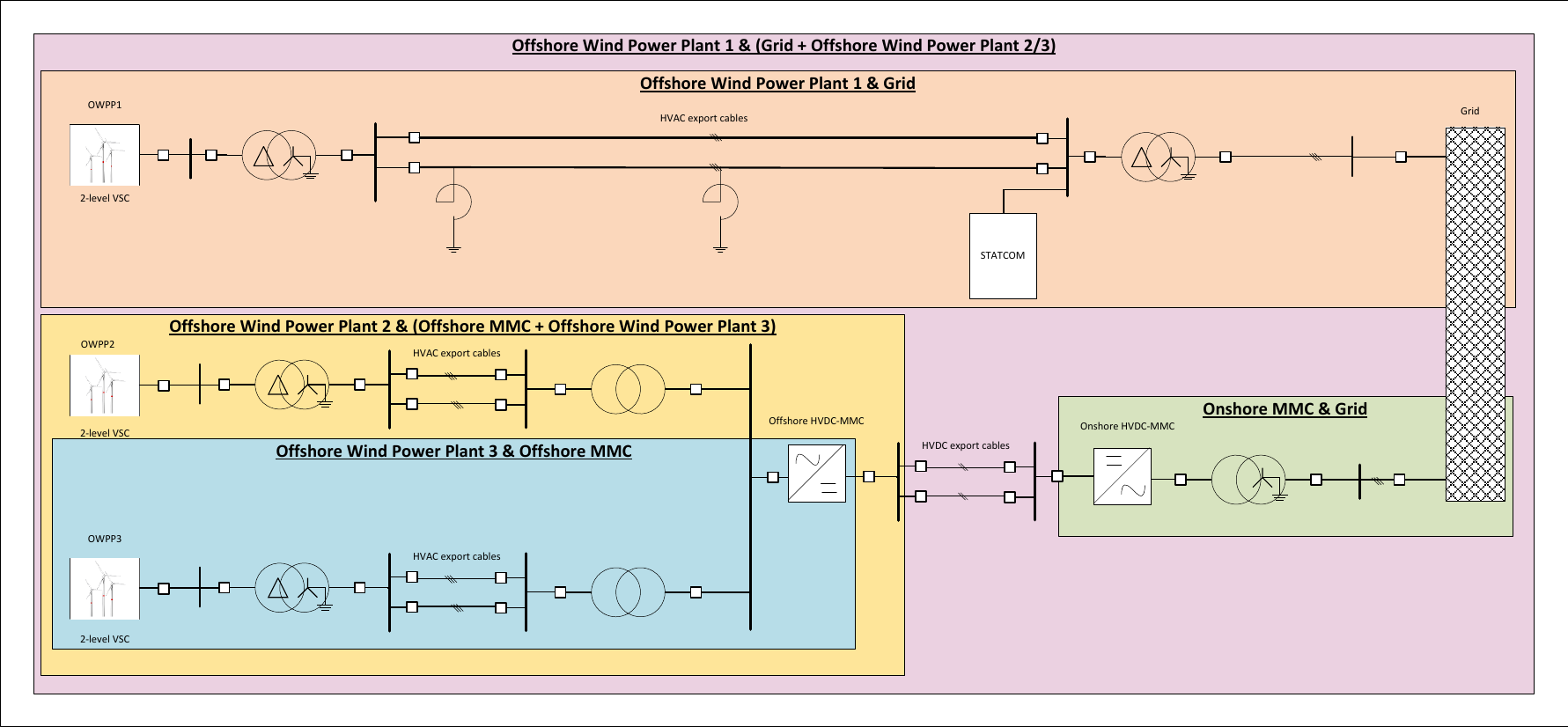}
    \caption{\ac{OWPP} Case Studies}
    \label{fig:casestudies}
\end{figure*}

The analysis consists of deriving a system's equivalent continuous frequency response at its point of connection via repetitive injection of voltages/currents at a defined amplitude for a range of frequencies. The response is then measured at the relative frequency, and output is a continuous input/output relation in impedance or admittance. The equivalent impedance can be derived analytically by obtaining Thevenin equivalents or more frequently via the frequency scanning technique \cite{Wang2017FrequencySystem}. Stability studies via the impedance-based method require splitting the system into two subsystems, i.e., source subsystem (A) and load subsystem (B); then, the ratio of the source impedance to load impedance must meet the Nyquist stability criterion for the system to be stable.
Typical cases, represented in Fig. \ref{fig:casestudies}:
\begin{enumerate}[label=\arabic*.,ref=\arabic*]
    \item Offshore Wind Power Plant \& Grid \label{case:1}\cite{Ghosh2019ImpedanceModel}
    \item Offshore Wind Turbines \& Offshore \ac{MMC} \label{case:2}\cite{Lyu2016FrequencyIntegration}
    \item Onshore \ac{MMC} \& Grid \label{case:3}\cite{Man2020Frequency-CouplingSystem}
    \item Offshore Wind Turbines Vendor A \& (Offshore \ac{MMC} + Offshore Wind Turbines Vendor B) \label{case:4} \cite{Zhou2021ASystems}
    \item Offshore Wind Power Plant A \& (Grid + Offshore Wind Power Plant B) \label{case:5} \cite{Jacobs2023ElectricOnline}
\end{enumerate}

The relationship between voltage and current is synthesized in a closed-loop transfer function $G_{CL}(s)$ given by equation \eqref{eqn:closeG}. The loop gain of the system transfer function is just the ratio between the equivalent impedance of both systems as shown in equation \eqref{eqn:openG}. Therefore, it is possible to analyze closed-loop stability by looking at the open-loop gain $G_{OL}(s)$, composed of just the equivalent impedances.  
\begin{equation}
    G_{CL}(s) = \frac{Z_A}{1+Z_A/Z_B}
    \label{eqn:closeG}
\end{equation}

\begin{equation}
    G_{OL}(s) = \frac{Z_A}{Z_B}
    \label{eqn:openG}
\end{equation}

Where $Z_A$ and $Z_B$ are the impedance-equivalent of the source and load, respectively.

The stability/system dynamics is then studied by analyzing the Bode Plot or Nyquist Plot \cite{Liao2018GeneralAnalysis}. On the Bode Plot, the analysis is done at the crossover frequencies where the two systems' equivalent impedance intersects as shown in equation \eqref{eqn:cutfreq}.
\begin{equation}
    \omega_c : |G_{OL}(\omega_c)|_{dB} = 0 \wedge |G_{OL}(\omega_c)| = \frac{Z_A}{Z_B} = 1
    \label{eqn:cutfreq}
\end{equation}
Once the intersecting frequencies are found, the Phase Margin is analyzed, representing the distance from a potentially unstable behavior as indicated in equation \eqref{eqn:PM}.
\begin{equation}
    PM = 180^{\circ} - \angle G(j\omega_c) = 180^{\circ}- (\phi_A - \phi_B) > 0
    \label{eqn:PM}
\end{equation}

Where $\phi_A$ and $\phi_B$ are, respectively, the impedance-equivalent phase of the source and load. The intersection frequency of source-load frequency-response is $\omega_c$.

Theoretically, the Bode Diagram cannot be used if the resultant system is unstable; therefore, the Nyquist Plot is used in parallel. In the Nyquist plot, it is possible to assess the stability by looking at the encirclement on the (-1,\ j0) point. When the Frequency Scanner is done via dq or sequence domain, the equivalent impedance is a 2x2 matrix that can be studied component-by-component with the Bode Diagram. Alternatively, the Generalized Nyquist Criterion can be applied, and the overall analysis can be done on the eigenvalues of the minor loop gain $L(s) = \frac{Z_A}{Z_B}$ at each frequency. The method's main advantage is that only the impedance of the two subsystems is needed; hence, when applied to black-box models, it could fournish a comprehensive understanding of the critical resonance frequencies.

Impedance analysis validation could be done via \ac{EMT} simulation. The resonance peaks identified via Impedance-based analysis are compared with the harmonic content in the signal via \ac{FFT} analysis \cite{Dhua2017HarmonicPlants}. The method limitations concern responsible state variable identification; thus, the response process is trial and error. Moreover, the Bode Diagram has to be provided for multiple operation points. While generally impedance-based studies are limited to the Phase Margin and Nyquist Stability, additional considerations could be done on the derivative of the frequency response since it may include meaningful information \cite{Liao2018GeneralAnalysis, Liao2020ImpedanceConverters}.

\subsubsection{Passivity-based}
Passivity-based analysis extends the impedance-based one to a broad frequency range rather than a specific resonant frequency. The concept is derived from Non-Linear Control Theory, where the main theorem states that the (negative) feedback connection of two passive systems is passive \cite{Khalil2014NonlinearSystems}. Passivity refers to a particular relationship of a component where input u and output y respect this law $u^Ty\geq0$. The link with the stability is that if a system is passive, it is also BIBO (bounded-input-bounded-output) stable. On the other hand, the opposite relation is not valid; thus, if a system is stable, it may not necessarily be passive. Consequently, if the system's components are passive, the overall interconnected system will also be passive, thus stable. It is possible to verify the passivity of a component via frequency-response $G(j\omega)$ (potentially derived via frequency scanning) by looking at these two relations visible in equation \eqref{eqn:pass1} and equation \eqref{eqn:pass2}:

\begin{equation}
    \angle G(j\omega) \in [-\frac{\pi}{2}, \frac{\pi}{2}]
    \label{eqn:pass1}
\end{equation}

\begin{equation}
    Re[G(j\omega)] \geq 0
    \label{eqn:pass2}
\end{equation}

When this is applied to a converter, thus the $G(j\omega)$ now is in the form of equivalent impedance $Z(j\omega)$, measured in dq-domain it will result in an impedance in \ac{MIMO} form where the converter is passive if equation \eqref{eqn:pass3} is satisfied. While for \ac{SISO}, the electrical element is passive if it satisfies equation \eqref{eqn:pass4}. In addition, the phase of output impedance $\angle Z(j\omega)$ must be within $[-\frac{\pi}{2}, \frac{\pi}{2}]$ and the converter must be individually stable \cite{WGC4.492024Multi-frequencyBROCHURE,Wu2023Passivity-BasedInverter}.
\begin{equation}
    \textbf{Z}(j\omega)+\textbf{Z}(j\omega)^{-1} > 0
    \label{eqn:pass3}
\end{equation}
\begin{equation}
    \text{Re}\{Z(j\omega)\} > 0
    \label{eqn:pass4}
\end{equation}

The passivity-based method looks efficient in guaranteeing the stability of two interconnected systems. However, its definition is generally applied to converters (not only in the context of \ac{OWPP}), which are not passive components since they are assumed to behave like current sources (\ac{GFL} control) or voltage sources (\ac{GFM} control) and they have constant power control \cite{Kocewiak2020OverviewSystems}. Hence, passivity cannot be guaranteed for the entire frequency range, particularly for frequency range close to fundamental frequency. In addition to the frequency range, the gain of the equivalent impedance should be high enough to guarantee passivity. However, high gains bring potential issues connected with abrupt responses and higher sensibility to disturbances, overshoots, and oscillations. This aspect might lead the converter control to saturate, thus falling into a non-linear zone. Consequently, the gain has to be designed accordingly to guarantee passivity while not decreasing the overall control's performance. 

Wu H. \textit{et al.} \cite{HubertAalborgMMC-HVDC} showed how guaranteeing the passivity of an \ac{OWPP} \ac{HVDC}, more specifically to the \acp{WT} and Offshore-\ac{MMC}, requires a balance in the controls. The active damping introduced in the controls improves the passivity in the low-frequency range. However, it will affect the passivity in the higher frequency zone. Therefore, the passivity requirement may not always be followed by the performance one. Control topology/schemes e.g. GFL/GFM, tuning of controls, additional damping, and filters, play a significant role in determining the passivity of a converter \cite{Wu2023Passivity-BasedInverter,Wang2022PassivityDamping,Akhavan2021PassivityVariations,Liao2020Passivity-basedConverters,Xie2020Passivity-BasedModel,Wu2020Virtual-Flux-BasedVSCs,Harnefors2017VSCAssessment}. Generally, passivity-based analysis has been motivated by the need to evaluate the extent to which the converter's equivalent impedance is non-passive. The approach is becoming more attractive compared to the impedance-based because the impedance-based approach needs to assess the stability when the network impedance varies repeatedly. Since the network impedance varies in a wide range in practice, passivity-based could be better in assessing the stability \cite{Harnefors2016Passivity-BasedOverview,Harnefors2008Frequency-domainDesign}. It is assumed that passive converter impedance will not cause stability issues in the grid. The method has recently attracted attention in industry and academia, but it's not yet fully mature because it is unclear how to apply it in the low-frequency range \cite{WGC4.492024Multi-frequencyBROCHURE}. 

\subsection{Eigenvalue/Modal Analysis Method}

 The Eigenvalue and Modal analysis are methods belonging to modern control systems, specifically to systems represented in state space form \cite{Dorf2017ModernEdition,P.Kundur1994PowerControl,M.J.Gibbard2015Small-signalSystems}. 
 Compared with the impedance or transfer-function methods, where poles (eigenvalues) are also used as stability indicators, the State Space representation can capture additional modes not visible in the former two due to zero-pole cancellations.   
\begin{equation}
\begin{cases} 
    \Delta \Dot{\textbf{x}} = \textbf{A} \Delta \textbf{x} + \textbf{B} \Delta \textbf{u} \\
     \Delta \textbf{y} = \textbf{C} \Delta \textbf{x} + \textbf{D} \Delta \textbf{u} 
\end{cases}
\label{eqn:ss}
\end{equation}

Where the \textbf{A} is the system matrix and $\Delta{\textbf{x}}$ is the state vector. 

Meaningful information can be collected by analyzing the system matrix and computing its associated eigenvalues and eigenvectors. The relative eigenvalues $\lambda$ are analyzed in the complex plane where it is possible to extract the attenuation $\sigma$ which measures the rate of decay or growth of a mode, natural/resonance frequency $\omega$ of the mode and relative damping/damping ratio $\zeta$  as illustrated in equation \eqref{eqn:eigen} and equation \eqref{eqn:damp}. The eigenvalues represent oscillatory and non-oscillatory modes of the system. For stability analysis, the oscillatory modes are of key interest.
\begin{equation}
    \lambda = \sigma \pm j \omega
    \label{eqn:eigen}
\end{equation}

\begin{equation}
    \zeta = -  \frac{\sigma}{\sqrt{\sigma^2+\omega^2}}
    \label{eqn:damp}
\end{equation}
The eigenvalues of the state matrix A are used to assess the system stability as follows \cite{P.Kundur1994PowerControl,M.J.Gibbard2015Small-signalSystems}:

\begin{enumerate}[label=\roman*.]
    \item The system is stable if all eigenvalues of state matrix \textbf{A} have a negative real part i.e. the damping ratio $\zeta$ is positive.
    \item The system is unstable if one or more of the eigenvalues of matrix \textbf{A} have a positive real part.
\end{enumerate}

The assessment of the stability above is only valid around the operating point. Special tools and techniques are required to convert the converter control system and grid model into a linear time-invariant (\ac{LTI}) system for converter-based power systems \cite{MatsLarsson2021SystematicAssessment}. The eigenvalues contribution could also be represented analytically if the \textbf{A}  matrix is not numerical but parametric. However, having an easily usable analytical equation is complicated if the state space has more than five states.  

Per each eigenvalue (or mode), it is possible to map each state's contribution via the participation factor analysis and act specifically on the main contributors \cite{Kocewiak2022PracticalMitigation}. The method consists of computing the right eigenvectors $\mathbf{\Phi}$ and left eigenvectors $\mathbf{\Psi}$ associated to \textbf{A}, computed with equation \eqref{eqn:rightvec} and equation \eqref{eqn:leftvec}.

\begin{equation}
    \mathbf{A}\mathbf{\Phi} = \mathbf{\Phi}\mathbf{\Lambda}
    \label{eqn:rightvec}
\end{equation}

\begin{equation}
    \mathbf{A}^T\mathbf{\Psi} = \mathbf{\Psi}\mathbf{\Lambda}
    \label{eqn:leftvec}
\end{equation}

Where $\mathbf{\Phi}$ and $\mathbf{\Psi}$ are matrices composed by equation \eqref{eqn:rightelements} and equation \eqref{eqn:leftelements}, which each element is related to a specific eigenvalue $\lambda_i$, collected on the diagonal of $\mathbf{\Lambda}$ matrix as shown in equation \eqref{eqn:eigenmatrix}.

\begin{equation}
    \mathbf{\Phi} = [\mathbf{\Phi}_1, \mathbf{\Phi}_2, ..., \mathbf{\Phi}_n ]
    \label{eqn:rightelements}
\end{equation}

\begin{equation}
    \mathbf{\Psi} = [\mathbf{\Psi}_1^T, \mathbf{\Psi}_2^T, ..., \mathbf{\Psi}_n^T ]^T
    \label{eqn:leftelements}
\end{equation}

\begin{equation}
\mathbf{\Lambda} = \begin{bmatrix}
\lambda_1 & ... & 0 \\
... & ... & ...\\
0 & ... & \lambda_n
\end{bmatrix}
\label{eqn:eigenmatrix}
\end{equation}

Finally, each column of the participation factor matrix is computed via equation \eqref{eqn:Pcompute} and composed as equation \eqref{eqn:Pcomposit}.

\begin{equation}
\mathbf{P}_i = \begin{bmatrix}
p_{1i} \\
p_{2i}\\
\vdots \\
p_{ni}
\end{bmatrix} =
\begin{bmatrix}
\phi_{1i}\psi_{i1} \\
\phi_{2i}\psi_{i2}\\
\vdots \\
\phi_{ni}\psi_{n1}
\end{bmatrix}
\label{eqn:Pcompute}
\end{equation}

\begin{equation}
    \mathbf{P} = [\mathbf{P}_1, \mathbf{P}_1, \cdots, \mathbf{P}_n ]
    \label{eqn:Pcomposit}
\end{equation}
Once the participation factor \textbf{P} matrix is composed, the modal analysis is performed as follows:

\begin{enumerate}[label=\roman*.]

\item Identify which are the eigenvalues $\lambda_i$ with the smallest damping $\zeta$.

\item In the participation matrix \textbf{P} identify per each undamped mode (column) which are the states $x_i$ that contribute the most (rows). 

\item Excite in a time-domain simulation the critical eigenvalue $\lambda_i$ by imposing as initial conditions $\mathbf{x}_0 = \mathbf{\Phi}_i$. This will show which are the states that contribute the most in time and identify the components/variables that oscillate against each other.

\end{enumerate}

In addition, it is possible to represent the right eigenvectors $\mathbf{\Phi_i}$ associated with the \textbf{A} matrix, which will show which variables (or devices) are oscillating against each other. The potentiality of the method consists in the possibility of collecting meaningful global information, more specifically, of each variable and its effect.

In general \ac{VSC} converters (for Type-4 \acp{WT} for example) are accurate enough and comparable to \ac{EMT} models \cite{Amin2017Small-SignalMethods}. The main concern are the number of the states that may lead to high computational effort. For example Kroutikova N. \textit{et al.} in \cite{KroutikovaState-SpaceMode} uses 17 states to represent a single converter that may be used to model generic \acp{VSC} (e.g. \acp{WT}, \acp{STATCOM}, simple \acp{MMC}, etc). While Yang D. and Wang W. designed a (indirect) grid-connected model with 21 states \cite{Yang2020UnifiedConverters}. Since \acp{OWPP} are composed by hundreds of \acp{WT}, instead of modelling each turbine it is possible to aggregate them \cite{Ghosh2019ImpedanceModel,Kunjumuhammed2017TheSystems}. For the future case where \acp{OWPP} will be composed by multi vendors, just a single state space model per vendor would be enough \cite{Liao2023StabilitySystems,ENTSO-E2021WorkstreamDevices}.

\subsection{Time domain simulations}
Time domain simulations are classified into two major types, i.e. \ac{RMS} and \ac{EMT} simulations \cite{Conseilinternationaldesgrandsreseauxelectriques.JointworkinggroupC4-C6352018ModellingStudies}. \ac{RMS}-based methods analyze the linearized components at a specific frequency, which allows the use of phasors to eliminate the time component from the steady state (linear) equations and linear algebra to solve the system. The \ac{RMS} simulations use large time steps which make it possible to simulate large power systems. Because of their large time steps, the analysis is possible in a limited frequency range. 

On the other hand, the \ac{EMT} is based on numerical solvers for differential equations,  %\eqref{enq:diffeqn}
% solved by using numerical methods such as Euler Method or Runge-Kuta as shown in equation \eqref{eqn:diffsol}. 
working with relatively smaller time steps (in order of microseconds) and hence simulation over a wider frequency range is possible. Since the converter dynamics extend beyond the fundamental frequency, \ac{EMT} tools become the most suited for multi-frequency stability analysis of converter-dominated systems and large-signal stability. Furthermore, non-linearities in the system are considered during \ac{EMT} simulations. \ac{EMT} simulations are considered reliable methods to validate other findings from small-signal stability analysis methods, giving engineers and researchers high confidence in their results, always considering that solution accuracy is dependent on the timestep.

Typical practice is to select feasible initial conditions derived from power flow solutions solved via the \ac{RMS} method in order to initialize the system correctly. The accuracy of the solution depends on the size of the solver's timestep as in equation \eqref{eqn:PSCADstep}.

\begin{equation}
    \Delta t = \frac{1}{10} \cdot \frac{1}{f_{\text{highest}}} 
    \label{eqn:PSCADstep}
    \end{equation}

The selection of $f_{highest}$ depends on the phenomena that are wanted to be included in the results, from ${10}^{-3}s$ for oscillatory transients or switching to ${10}^3s$ for voltage fluctuations. In the case of MATLAB/Simulink\textsuperscript{\textcopyright}, selecting from fixed or variable timestep allows for computational efficiency when the solution behavior is not steep. However, the simulation time is one of the primary limits, as accurate solutions require smaller timesteps, thus longer simulation time. In addition, compared to the Eigenvalue analysis method, \ac{EMT} simulations are not able to pinpoint the component or set of devices causing instability in the system \cite{ENTSO-E2021WorkstreamDevices}. 

The accuracy of the Time domain method for stability studies also depends on the number of cases considered, and in general, it is not useful for large systems \cite{Watson2020AnTechnology}. Due to the long simulation time, a limited set is always considered if a large number of study cases are selected, which risks the non-inclusion of contingency that triggers instability. Another disadvantage of using \ac{EMT} tools is that when different vendors are involved, the models have different time steps and may be software-specific compatible.

However, \ac{EMT} is indispensable to validate the cases analyzed via more analytical methods. Therefore, a combination of analytical methods, \ac{EMT} and real-time simulation is privileged in the literature. 
Besides, with \ac{EMT} alongside the generic or ad-hoc models, it is possible to share compiled Black-Box models via Dynamic Link Library (DLL) in order to guarantee the \ac{IP} protection \cite{VordemBergeSOLUTIONSSYSTEMS}.

The \ac{EMT} models are based on some simplification that might not capture all the valuable information. Therefore, to obtain the behaviour of the components in a real-time environment, real-time simulation using the hardware is used. Either of the following can perform a stability analysis of converter-based systems, including multi-vendor via real-time simulations: 

\subsubsection{Hardware-in-the loop  and Replicas}
Hardware-in-the-loop (\ac{HIL}) simulation enables testing of component's response in a real-time environment using their hardware. The control hardware is connected in the loop to evaluate how the actual devices will work on-site. The \ac{HIL} has gained popularity in industry and academia as it provides confidence in the controllers before deployment. When controllers from different manufacturers are interconnected, the \ac{IP} has to be preserved. To de-risk their control and protection systems, an approach recommended by the BestPaths project has gained popularity \cite{P.Rault2018D9.3:Systems}. The approach is for the vendors to provide detailed, black-boxed replicas of their control and protection hardware and \ac{EMT} models to a third party that carries out offline \ac{EMT} simulations and \ac{HIL} real-time simulations using replicas. This approach has been used in the Johan Sverdrup project to investigate control interactions between two \ac{VSC}-\ac{HVDC} converters from different vendors operating in parallel to supply an offshore oil field \cite{RTDSTechnologies2022ENSURINGTESTING}.

Although the approach captures all possible stability issues for the selected contingencies, it is unsuitable for offshore grids with many converters because of the many possible configurations to analyze \cite{VordemBergeSOLUTIONSSYSTEMS}. Additionally, an Iterative process is required, and it is time-consuming and expensive. Setting up the test facility will be challenging if multiple vendors are involved \cite{VordemBergeSOLUTIONSSYSTEMS}. The \ac{HIL} simulation was also used in \cite{Mitra2014OffshoreStudy} to test \ac{HVDC} \ac{OWPP} connected to the weak grid in different conditions and faults.

\subsubsection{\ac{EMT} co-simulation of multi-vendor models}
Verification of the dynamic performance of converter control and protection systems before site delivery using either online or offline \ac{EMT} tools is a requirement for manufacturers. Co-simulation involves using \ac{EMT} tools to co-simulate \ac{EMT} offline models from different vendors using separate computers linked via fast communication schemes. The models could be in the same or different physical locations. This enables control interaction analysis while safeguarding the intellectual property of the vendors \cite{O.D.ADEUYI2020Multi-terminalSchemes}. This approach is useful in analyzing the effect of integrating new converters in an existing network that has converters from other vendors.

\subsubsection{Frequency response determination using frequency scanning/sweep}
With the upcoming multi-vendor offshore wind infrastructure, using \ac{EMT} tools to obtain frequency response models from vendor black box models will be inevitable. This is a powerful approach as the response from the black box models can be checked to see if it matches the vendor-provided small-signal models.

When a system is too complex or only black-boxed models are provided, it is possible to use the frequency scanning technique to obtain equivalent impedance/admittance of the system/model using \ac{EMT} tools or on real systems/devices. Frequency scanning is derived from the frequency response theory where a \ac{LTI} system is excited via a signal in a sine waveform showed in equation \eqref{eqn:input} and the output is a sine waveform with potentially a different amplitude and phase $\phi$ but with the same frequency $\omega$ as the input signal. The output then is a sine-wave at the same frequency, but with a gain $H(j\omega)$ and a phase shift $\angle H(j\omega)$ produced by the system visible in equation \eqref{eqn:output}. 

\begin{equation}
    u(t) = A cos(\omega t + \phi)
    \label{eqn:input}
\end{equation}

\begin{equation}
    y(t) = |H(j\omega)| A cos(\omega t+\phi + \angle H(j\omega))
    \label{eqn:output}
\end{equation}

The excitation/disturbance signal can be injected with a voltage in series with the system or by injecting a current in parallel. According to Shirinzad M. \cite{Shirinzad2021FrequencySystems}, the voltage injection in ref{fig:voltageinj} should be applied on devices with a current source predominantly behavior. On the other hand, voltage source devices should be scanned by injecting current as in ref{fig:currentinj}. In any case, response validity is strictly correlated with \ac{LTI} systems, \cite{Rik2012SystemApproach}, thus linear operational points.

\begin{figure}[h]
    \centering
    \includegraphics[width=1\linewidth]{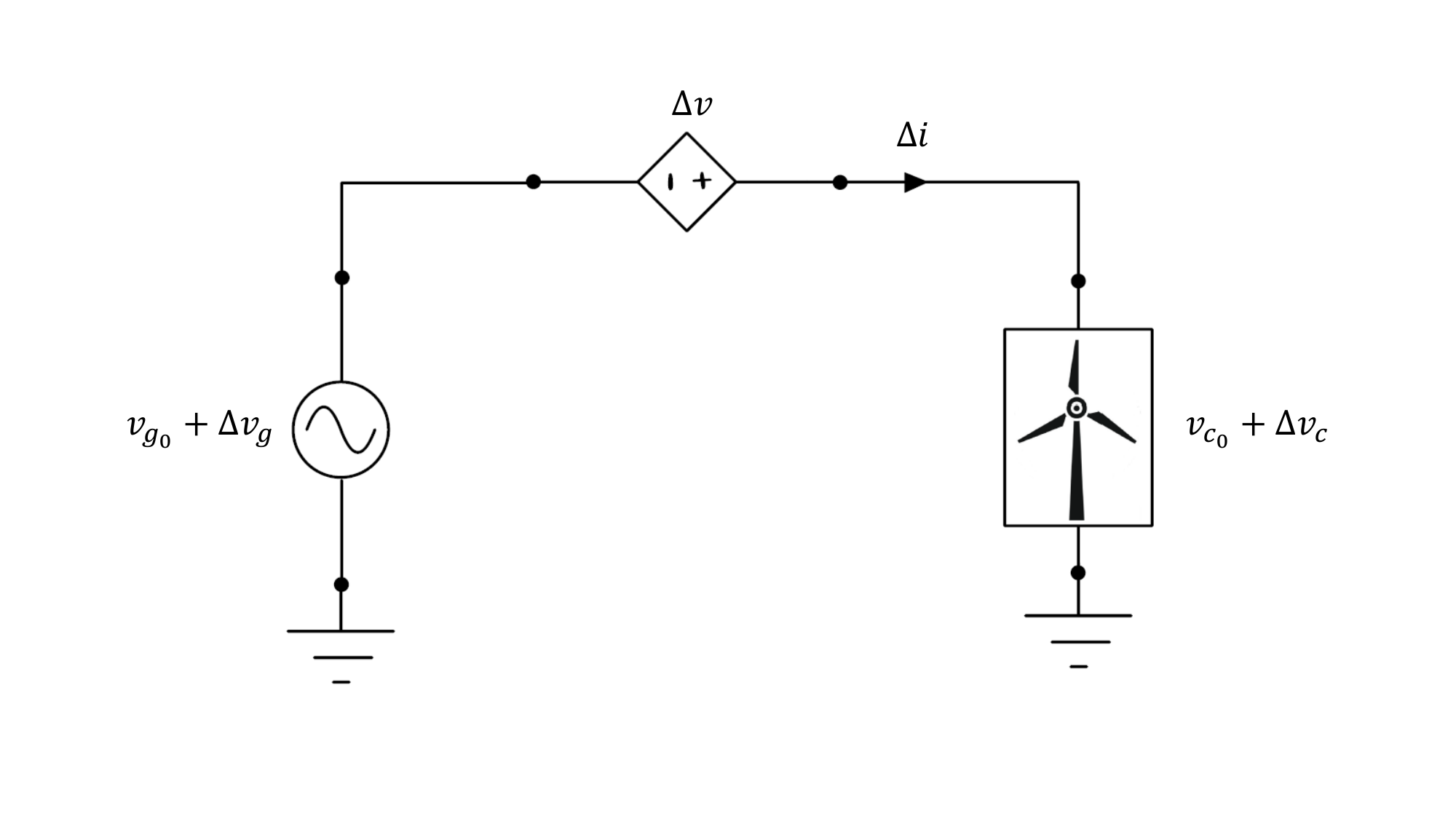}
    \caption{Voltage injection}
    \label{fig:voltageinj}
\end{figure}

\begin{figure}[h]
    \centering
    \includegraphics[width=1\linewidth]{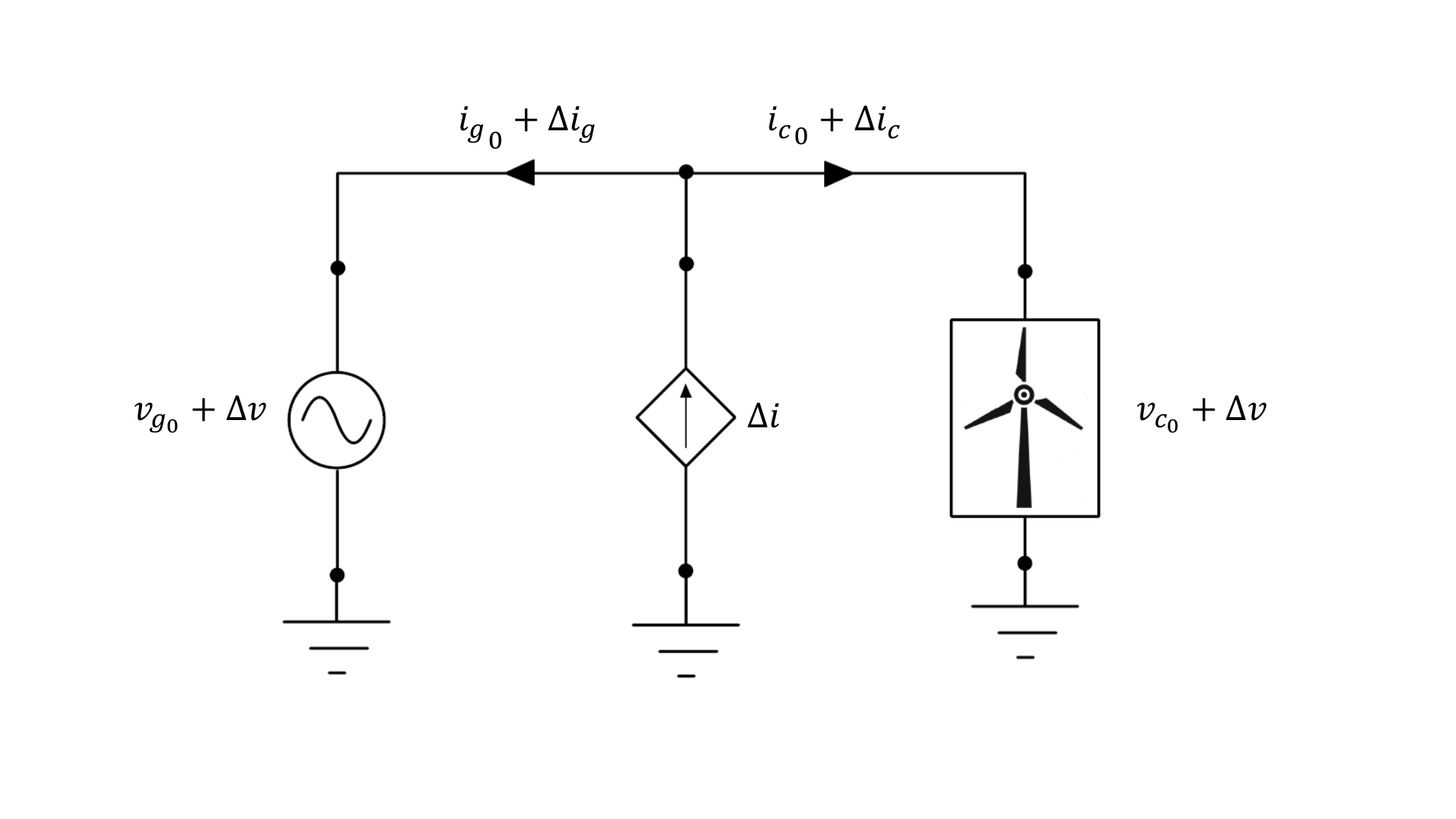}
    \caption{Current injection}
    \label{fig:currentinj}
\end{figure}

\paragraph{Single-Tone and Multi-Tone injections}\mbox{}\\
With single-tone injection, the system is disturbed by a single-frequency signal each time until all the desired frequency range is analyzed. At the same time, for multi-tone, perturbation is done by a wide-band signal. The single-tone is more accurate as it offers the highest signal-to-noise ratio but is time-consuming because a new simulation is needed for each perturbation frequency. Although it is time-consuming, the approach is preferred because frequency couplings that may be present in the impedance/admittance models of the components can be identified.

On the other hand, the multi-tone is faster, but because of multiple injections, the maximum amplitude of the perturbation signal is relatively small compared to that of a single-tone. The small amplitude might lead to a small signal-to-noise ratio, which in turn can lead to inaccurate results. While a larger magnitude of multi-tone injections can solve the problem of a small signal-to-noise ratio, it introduces spikes in the output response, which may disturb the system's operating point. Hence, the choice of perturbation signal should balance the accuracy of the scanning results with the simulation time \cite{Rygg2018Impedance-basedSystems,Rygg2016ASystems,DasComparisonSystems,Jiang1995ASYSTEMS,Nouri2021TestPerturbations}. It is recommended that the amplitude of the perturbation be kept below $5\%$ of the nominal values. The frequency of interest component should be identified in both voltage and current \ac{FFT}. Otherwise, a larger amplitude of the perturbation is needed as the signal-to-noise ratio is small.  The frequency of the perturbation signal should cover the entire frequency range and should have a sufficient frequency resolution.

The injected phase of the sinewave has to be uniformly spread in $\phi\in\left[0,\pi\right]$ not to overlap and produce a signal with an amplitude able to move the system to a critical point. 
The M. Shirinzad's multi-tone injection is a sum of shifted sinewaves illustrated in equation \eqref{eqn:multisin} with the same amplitude $a$, set frequency interval $f_d$ and a shift $\delta_l$ selected according to Schröder phases showed in equation \eqref{eqn:Schröder}  \cite{Pintelon2012SystemApproach}  (method used also in electrical machines field \cite{KimpianMultiphasePractice}). Similarly, H. Gong \textit{et al.} \cite{Gong2018ParametricConverters} used the \ac{PRBS} that injects binary-generated square waves.

\begin{equation}
    u(t) = a \sum^N_{l=l_0} sin(2\pi f_d l t+ \delta_l)
    \label{eqn:multisin}
\end{equation}

\begin{equation}
    \delta_l = - \frac{(l-l_o)(l-l_o+1)}{N-l_0+1}\pi
    \label{eqn:Schröder}
\end{equation}

Although the multi-tone injection is less accurate than the single-tone, it can find good applications in converter or grid online impedance estimation. The possibility to estimate the impedance with a single measurement on real systems ensures that the output is relative to a precise operational point \cite{Khan2024EnhancedSequence}.

\paragraph{Analysis via Bode and Nyquist diagrams} \mbox{}\\
Interpretation of the result could be done via Bode or Nyquist diagrams. The construction of both diagrams follows this pattern:

\begin{enumerate}[label=\roman*.]
    \item Inject a signal $u_i$ at frequency $\omega_i$
    \item Measure the relative output $y_i$
    \item Calculate the ratio $|H\left(j\omega\right)|$ between output $y_i$ amplitude and the input $u_i$ one 
    \item Calculate the difference in phase $\angle H\left(j\omega\right)=\angle y-\angle u$
    \item Repeat the procedure per each frequency $\omega_i$
\end{enumerate}
The Bode diagram is then constructed by plotting the magnitude and phase separately on magnitude-frequency and phase-frequency plots. The Nyquist plot is constructed by plotting on a complex plane $\bar{H}\left(j\omega\right)=\left|H\left(j\omega\right)\right|\angle H(j\omega)$ where the frequency is implicit. 

\paragraph{\ac{MIMO} analysis}\mbox{}\\
When the study is performed on a three-phase device or system depending on the signal injection and processing techniques, there is the possibility of using different reference frames, such as the Phase Domain, $\alpha\beta$-Domain, the Sequence Domain, and dq-Domain \cite{DasComparisonSystems}. Jacobs K \textit{et al.} \cite{Jacobs2023AParks} have tested the different scanning methods on a power system with WPP; the sequence domain is the fastest method, and  Shirinzad M. \cite{Shirinzad2021FrequencySystems} proved that it is more suitable for passive elements. If the scanning is performed on converter-based devices, the result is susceptible to the frequency mirroring effect when the injection is done in phase or sequence domain. This effect must be taken into account when analyzing system response from the frequency scans otherwise, the results will be inaccurate \cite{Bakhshizadeh2016CouplingsConverters,Ren2016AFarms}. The dq-Domain is generally more accurate and not affected by the Frequency Mirroring Effect. 

The injection in dq-Domain preserves the system's overall \ac{LTI}, although the same system would not be any more \ac{LTI} with injection in Phase or Sequence Domain. There is the possibility to apply the Modified Sequence Domain to avoid the frequency coupling in this frame \cite{Rygg2016ASystems}. Therefore, the choice of reference frame will depend on the frequency range of interest, the characteristics of the component to be scanned, and the features of the \ac{EMT} model from the vendors \cite{WGC4.492024Multi-frequencyBROCHURE,Rik2012SystemApproach}. 

When the scanning is performed in dq-Domain, the computational effort increases since the scan will not be anymore a \ac{SISO} relationship but a \ac{MIMO} one. The general steps recommended by CIGRE for the scanning procedure are \cite{WGC4.492024Multi-frequencyBROCHURE}:

\begin{enumerate}[label=\roman*.]
    \item Initialize the system at steady state in a \ac{EMT} simulation
    \item Inject a voltage/current at a specific frequency $\omega_d$ in d\&q. The sum of the steady state contribution and the injected one will have a form visible equation \eqref{eqn:dqinjection}. In order to have the $V_{d0/q0}$ as a constant contribution, the angle $\theta$ reference for the Park transformation should be $\theta = \omega_n t$, where $\omega_n$ is the nominal frequency of the system under study. The magnitude of the injection $V_d$ should be large enough not to be confounded with the noise and small enough not to move the system to a different operational point or saturate the controllers in case of \acp{IBR}.

    \begin{equation}
        V_{d/q}(t) = V_{d0/q0} + V_icos(\omega_i t), V_{q/d} = 0
        \label{eqn:dqinjection}
    \end{equation}
    \item  The current is measured in \textit{d} ($I_d$) and \textit{q} ($I_q$) frame
    \item The injection is repeated for the other voltage component
    \item The relative admitance $\textbf{Y}$ is computed via equation \eqref{eqn:Ydq}.
    \begin{equation}
    \left\{
    \begin{array}{l}
    Y_{dd}(j\omega_i) = \frac{I_d}{V_d} \\
    Y_{dq}(j\omega_i) = \frac{I_d}{V_q} \\
    Y_{qd}(j\omega_i) = \frac{I_q}{V_d} \\
    Y_{qq}(j\omega_i) = \frac{I_q}{V_q}
    \end{array}
    \right.
    \label{eqn:Ydq}
    \end{equation}
    \item The procedure is repeated for the other frequencies in interest. 
    
\end{enumerate}
The resultant transfer functions will be four different input-output relations that are usually stored in the form of a 2x2 matrix as showed in equation \eqref{eqn:Zmatrix}.

\begin{equation}
    \overline{\textbf{Z}} = 
\begin{bmatrix}
    \overline{\textbf{Z}}_{dd}  & \overline{\textbf{Z}}_{dq} \\
    \overline{\textbf{Z}}_{qd}  & \overline{\textbf{Z}}_{qq}
\end{bmatrix}
\label{eqn:Zmatrix}
\end{equation}
To study the results in this format it is possible either to analyze separately the Bode \& Nyquist diagram of each component. Otherwise, it is possible to apply the Generalized Nyquist Criterion by studying the Bode \& Nyquist diagrams of the equation \eqref{eqn:Zmatrix} eigenvalues per each frequency.

Shirinzad \cite{Shirinzad2021FrequencySystems} claims that the non-time invariant switching converter causes the Frequency Mirroring effect while Liao Y. \cite{Liao2020ImpedanceConverters} noted that in \ac{PLL}-based converters the phenomenon could be caused by the asymmetry of the \ac{PLL}'s controller where just the q-component is controlled. In converter-based systems, the injection on the \ac{AC} side is generally reflected also on the \ac{DC} side, where the constant component will be added to an oscillatory one. The result is that the converter will no longer modulate a constant \ac{DC} voltage but a time-varying sine with a \ac{DC} offset. If the output at \ac{AC} is decomposed via the Fourier Series, the Fourier coefficients are no longer constant but harmonically varying. This is translated as additional harmonics, visible when the Double Fourier Series is performed, around the modulating one \cite{PulseXplore}. 

In the context of offshore wind, the frequency scanner could be used by:
\begin{enumerate}[label=\roman*.]
    \item \ac{TSO}: for pre-screening stability risk when new converters are connected.
\item \ac{WT} Manufacturers: to ensure that the product meets dynamic specifications at \ac{PoC}.
\item \ac{OWPP} Developers: to assess the stability of different interconnected components.
\end{enumerate}
Frequency scans can also be used to validate that the impedance provided by the manufacturers matches their \ac{EMT} models. The main advantages of this approach are:

\begin{enumerate}
    \item Using \ac{EMT} tools, it is possible to obtain converter equivalent impedance over its controller bandwidth.
    \item It is possible to obtain a state space model or transfer function of the system by utilizing the vector fitting technique.   
    \item Equivalent impedance derived via frequency scan can be used for impedance-based stability analysis.
\end{enumerate}

\section{Discussion and future research outlook}
\label{sec:discussion}
% Intro
The decarbonization agenda has led to the paradigm shift from traditional sources to renewable intermittent sources, which are converter-based. The shift has seen exponential growth in \ac{OWPP}. The current power system was not designed to host these relatively new types of power sources; therefore, the focus on the system is shifting from power quality to stability issues. The shift is also being experienced in \ac{OWPP} topologies with advocacy to change from the current single vendor point-to-point \ac{HVAC} or \ac{HVDC} to the multi-vendor \ac{OWPP} and multi-terminal \ac{HVDC} to enhance controllability and redundancy.

The new interconnection presents challenges to both system operation and control, and hence, stability analysis is key to addressing the challenges presented. Several methods have been discussed in this paper. The analysis methods are classified into frequency domain and time domain, each with its limitations depending on the type of investigation. The methods have been used extensively in academia but are also gaining popularity in the industry, especially by \acp{TSO} and manufacturers. Generally, when the stability analysis methods are applied by considering \acp{OWPP} from the \ac{PoC}, they are applied without considering the complexity behind it since \acp{OWPP} is characterized by a complex transmission system within the power plant itself. The procedure to perform stability analysis is to first screen the system for problematic scenarios and then do more detailed studies to quantify the scenarios that present undesirable responses. 

When \ac{WT} manufacturers and \ac{OWPP} developers design and tune the parameters of a plant, the condition of the grid at the \ac{PoC} where the \ac{OWPP} will be installed is necessary. However, with the future penetration of more \ac{IBR} in the power system, traditional system screening methods are proven to be no longer trustworthy of the actual grid's conditions at the \ac{PoC}.

% System screening
System screening is done mainly by two approaches, i.e., determining the interaction factors and system strength. These two metrics are important for the initial assessment of the system. Interaction factors measure voltage sensitivity between two buses and are useful when new plants are integrated into the system. A factor of less than 0.15 indicates a high possibility of interaction between the components connected in the two buses. The \ac{OWPP} developers mostly use the metric when new \acp{OWPP} are integrated into the onshore grid. However, the metric may not give a correct view of interaction possibilities in the context of \ac{OWPP} clusters that share connection points offshore. Since clusters will be connected to the same bus, a high value will be obtained, indicating fewer chances of interactions, but there is a very high possibility of control interaction between hundreds of converters, especially if the clusters are from different vendors. 

Traditionally, system strength has been assessed using \ac{SCR}, an indicator of the p.u.
impedance of the system at the fundamental frequency. A value of less than three shows that the system is weak and susceptible to instability, especially if the converter is in \ac{GFL} control mode. Since the \ac{SCR} considers only the physical line impedances, it is not valid to characterize a system dominated by converters. The new proposed metrics (\ac{GSIM} and \ac{IMR}) take into consideration the converter behavior across a frequency range. Although this is a promising development, the metrics are only at the research stage, and more validation tests are needed to have a confidence level equivalent to that of the traditional \ac{SCR} method for them to be adopted by the industry.

Incorrect system strength characterization means a new \ac{OWPP} will use the wrong metric to tune its controls. Specifically, when an \ac{HVAC}-\ac{OWPP} has to be connected, the controls of the single \acp{WT} must be tuned accordingly to the grid conditions. On the other hand, with an \ac{HVDC}-\ac{OWPP}, just the onshore \ac{MMC} has to be correctly tuned. Concerns might arise when in close proximity (in electrical terms) are placed other \acp{IBR}, namely other \acp{OWPP}. In this case, the subsequent plant has to be tuned not to interact with the other \ac{OWPP} in addition to respecting the requirements at the \ac{PoC}. At the same time, the introduction of multiple \acp{IBR} in an area affects the local system strength. Therefore, the condition at the \ac{PoC} for the first \ac{OWPP} are no longer the same; therefore, the tunings have to be changed accordingly. Hence, a wrong metric can be a catalyst for instability.

% Transfer function
The Transfer function method can be considered the forerunner of the other frequency-domain-based ones. However, its application for \acp{OWPP} is restricted due to the fact that accurate studies, even for a relatively simple point-to-point \ac{HVAC} structure, would need to take into account the multiple cascade \ac{VSC} controllers in series with components (filters, transformers, etc.), frequency-domain model of the \ac{HVAC} cable and the grid. This would result in an impracticable formulation. 
However, if a transfer function formulation is needed for specific applications, it is possible to derive it directly from the State-Space formulation. Therefore, it is not advisable to confront stability studies of \acp{OWPP} directly via transfer functions. 

% Impedance 
The impedance-based method applied to \acp{OWPP} is essentially a continuous frequency response of offshore system subsections. The method is extremely versatile and potentially breakthrough even in the last part of a \ac{OWPP}. The analysis could be applied to real devices or offshore \ac{EMT} black-box models. If the analysis is done offline via \ac{EMT} black-box models, it should perform better when applied to cases \ref{case:2} and \ref{case:4} of Figure \ref{fig:casestudies}. This is because when taking into account the offshore side of an \ac{HVDC}-\ac{OWPP}, the model is composed of accurate vendor models at both ends. After all, the onshore part could be considered fully decoupled. However, when studies are performed on cases \ref{case:1}, \ref{case:2}, and \ref{case:5}, it is necessary to have an accurate grid representation in the frequency domain; otherwise, the results might not be completely accurate. It is possible to use curves provided by the \ac{TSO} (if available). However, it is mandatory to ensure that the relative curves of the converters are built on the same operational conditions as the ones provided by the \ac{TSO}. In order to validate the impedance-based results, a good practice is to perform the \ac{EMT} study (on the same operational point as the impedance-based one) and analyze via \ac{FFT} if the resonance peaks match the ones visible in the Bode Diagram. 

% uses an index that does not have physical meaning in the context of \ac{IBR} generation. 
% Passivity
The passivity-based method is an extension of the impedance analysis. The main difference is that impedance-based needs the model for both grid and \ac{IBR}, while with passivity, only the \ac{IBR} should be checked for passivity. However, this is only valid if the \ac{TSO} also uses the passivity criterion.

While passivity can guarantee stability, it may influence performance, e.g., it may not guarantee grid code compliance. Nevertheless, \ac{HVDC}-\ac{OWPP} multi-vendor could still benefit by using this stability method for the offshore \ac{AC} system (decoupled from the onshore). Even if the offshore network has reduced performances, it does not have to submit to the same rules (in the current Grid Codes context), therefore guaranteeing the passivity of the \acp{WT}' and offshore-\ac{MMC} could still be a meaningful application. On the other hand, for the onshore-\ac{MMC} or \ac{HVAC}-\ac{OWPP}, this might be valid if the \ac{TSO} is also using the passivity as a criterion at the \acp{PoC}.

% Eigenvalues
The Eigenvalues/Modal analysis finds extensive applications in the literature, and it is also becoming attractive for \ac{WT} manufacturers and \ac{OWPP} developers due to its versatility. More specifically, it could find useful applications in the context of the \ac{HVDC}-\ac{OWPP} offshore side when converter models are provided in state-space form, where the system is fully converter-based. In the multi-vendor case, detecting the instability conditions via the participation factor might be possible, and responsibility can be assigned by spotting the specific variable. Generally, models are not available due to \ac{IP} protection. Therefore, extracting a state space form via Vector Fitting or sharing numerical state space (dependent on the operational point) might be possible. In addition, it introduces the possibility for sensitivity analysis via parameters sweep and tuning.

On the other hand, for what concerns \ac{HVAC}-\ac{OWPP} or onshore \ac{MMC} for \ac{HVDC}-\ac{OWPP} in the absence of a state-space model of the Grid, it might not be possible to study the interaction between the \ac{OWPP} and the power system. However, the method is still valuable for analyzing the interactions depending on other parameters within the \ac{OWPP} or systems connected in parallel as \acp{STATCOM} or other \ac{OWPP} in close proximity.

In order to validate the results or applied controls in this form, it is common to analyze the small-step response via \ac{EMT} simulations. In addition, it might be possible also to use impedance analysis to verify if the resonance behavior matches. 

%time domain
The \ac{EMT} simulation is the most accurate method that has been used to corroborate stability issues found by other stability analysis techniques. The confidence in the \ac{EMT} simulation can be further enhanced by hardware-in-loop real-time simulations. Because of larger computation requirements, study cases should be selected carefully to capture the intended scenarios as well us maintain reasonable computation time. The dynamic stability of \ac{OWPP} clusters can be accurately analyzed using the \ac{EMT} black box models provided by the vendors. By using the \ac{EMT} frequency scanning technique, impedances for black box models and grids can be measured, which can then be used to develop linear models for other small-signal stability analysis tools. The main advantage of obtaining the linear models from \ac{EMT} tools is that the impact of nonlinearities from the converter, transformer saturation, etc., are considered. 

The \ac{EMT} tools are applicable at any stage of \ac{OWPP} development; however, they are faced with the challenge of different vendors having specific timesteps for their models. In addition, the \ac{EMT} models provided are accurate only for the specific study cases. However, since \ac{RMS} methods are not sufficient to represent system dynamics for a converter-dominated \ac{OWPP} system, \ac{EMT} tools are still recommended. Since the simulation time is long due to small timesteps that make the approach tedious for large systems, it is important going forward to develop \ac{EMT} models that correctly represent the phenomenon of interest without much modeling complexity.

% Limits, suggestions and future work

\begin{figure}[h!]
    \centering
    \includegraphics[width=1\linewidth]{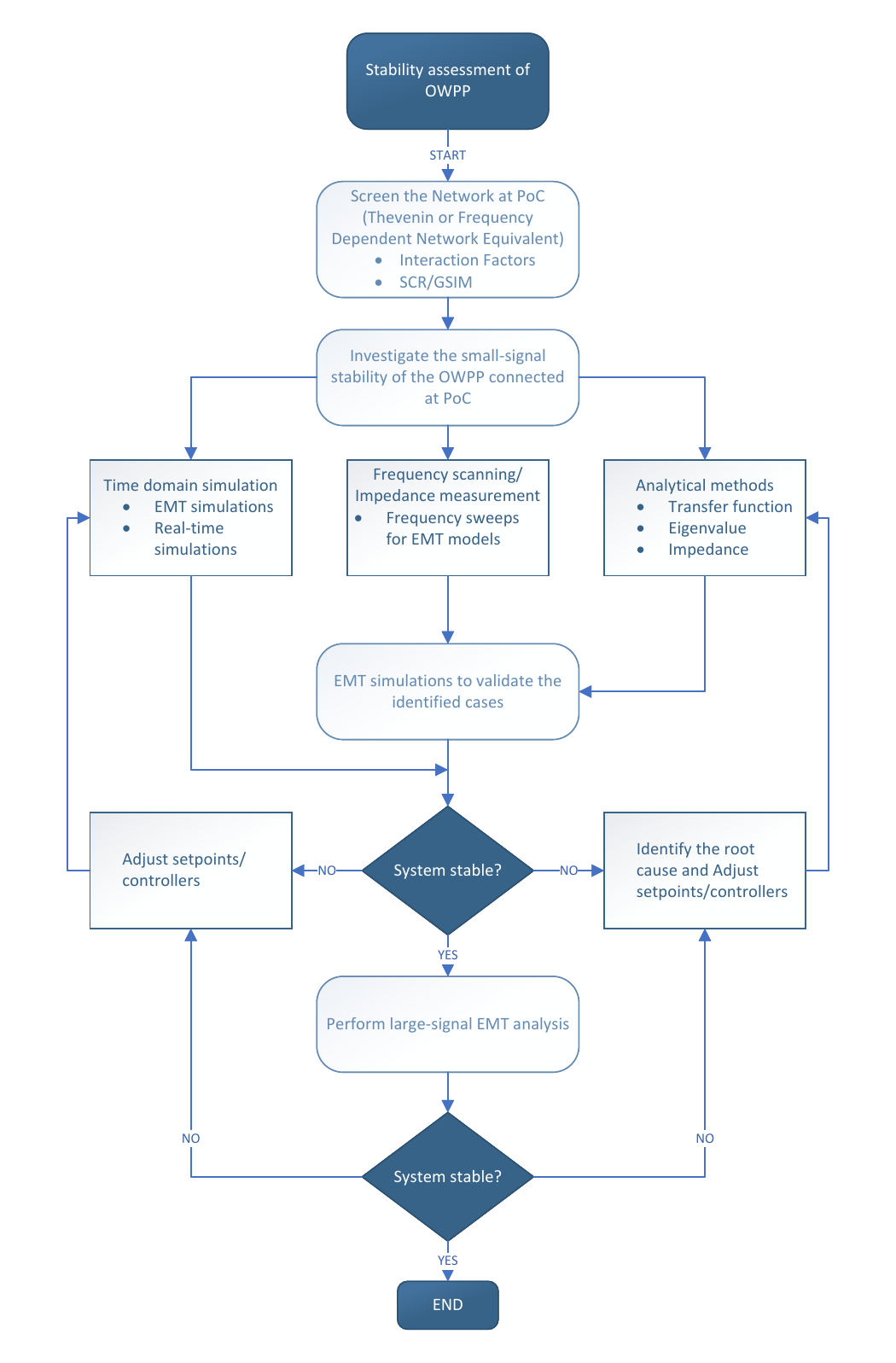}
    \caption{Stability analysis workflow for \ac{OWPP}}
    \label{fig:Stability_flow}
\end{figure}

Based on the above analysis, it is clear that each method has some limitations that may limit its application in certain cases, as summarized in \ref{tab:comparison_table1} for screening methods and \ref{tab:comparison_table2} for stability analysis methods. 
Thus, to correctly analyze the stability of the offshore system, a combination of the methods is needed to achieve the much-needed accuracy and trustworthiness of the analysis. 
In this era of converter-based systems, a method that captures converter dynamics is preferred. In this regard, the authors recommend obtaining linear models using the \ac{EMT} frequency scanning tool, which can then be used to obtain the system's frequency response. For example, impedance matrices obtained from the scans can be used to assess the stability of the system using the well-defined metrics of the impedance-based approach. Alternatively, the eigenvalues of the system can be calculated from the state-space model, allowing for identifying the instability root or the component contributing to it. A summary of the stability analysis workflow for \ac{OWPP} is proposed in Fig. \ref{fig:Stability_flow}.

To accelerate growth and guarantee the dynamic stability of large offshore wind clusters, there is a need to expedite research in the following areas:
\begin{enumerate}[label=\roman*.]
\item Development of dynamic models capable of accurately simulating large multi-vendor offshore wind power plants.
\item Detailed studies on how to analyze interoperability in multi-vendor \ac{OWPP} clusters. 
\item Standardization of metrics for assessing system strength for a converter-dominated system. 

\item Move towards partially open models for the upper converter control layers.

\item Design standard validated grey-box models for converters, which can be fine-tuned using \ac{EMT}, Vector fitting, or Impedance methods, to mimic the dynamic behavior of black-box components protected by \ac{IP}. 
\end{enumerate}

%% Comparison of the methods
\begin{table*}[h!]
\tiny
\caption{Comparison of system screening methods}
\resizebox{\textwidth}{!}{%
\centering
\begin{tabular}{>{\raggedright\arraybackslash}p{3.0cm}>{\raggedright\arraybackslash}p{3.96cm}>{\raggedright\arraybackslash}p{3.96cm}>{\raggedright\arraybackslash}p{3.96cm}}
\hline
\textbf{Method} & \textbf{Advantages} & \textbf{Limitations} & \textbf{Practical Application} \\
\hline
Interaction Factors  \cite{Conseilinternationaldesgrandsreseauxelectriques.ComitedetudesB4.2008SystemsInfeed,Henderson2024GridSystems} & \vspace{-4mm} \begin{itemize}[leftmargin=*] 
\item Useful for initial screening to identify potential instability scenarios \end{itemize} &\vspace{-4mm}\begin{itemize}[leftmargin=*]
\setlength\itemsep{-0.5em}
\item  Calculated within fundamental frequency; hence, control interactions across the controller bandwidth are not considered
\item It is not applicable for \ac{OWPP} clusters connected at the same bus. In this case, voltage sensitivity will be one, suggesting no possible interactions, but there could be potential interactions from the hundreds of converters, especially if the clusters are multi-vendor.
\item The metric is just a number that provides a general overview of system stability but cannot pinpoint the causes of instability \end{itemize} & It is used by \ac{OWPP} developers for initial multi-infeed screening study. The screening of \ac{OWPP} clusters to be connected to the same bus offshore is not feasible  \\
\hline
Short-circuit ratio (\ac{SCR})  \cite{WGC4.492024Multi-frequencyBROCHURE,Henderson2024GridSystems,Henderson2022AnalysisSystems,Henderson2022ExploringConverters,Conseilinternationaldesgrandsreseauxelectriques.ComitedetudesB4.2016ConnectionNetworks,NERC2017NERCSystems,Davari2017RobustDynamics}  &\vspace{-4mm} \begin{itemize}[leftmargin=*] 
\item Useful for initial screening to assess the system strength 
\item Well know with well-defined metrics to assess system strength \end{itemize} &\vspace{-4mm}\begin{itemize}[leftmargin=*] \setlength\itemsep{-0.5em} 
\item Calculated within fundamental frequency; hence, possible resonances in the grid are not considered 
\item It can lead to the wrong characterization of the converter-dominated system, such as \ac{OWPP} clusters, because the converter control impedance is not considered  
\item The information from the metric cannot be used for detailed quantitative study\end{itemize} & Has been used extensively and reliably to assess the strength of the traditional power system \\
\hline
Grid Strength Impedance Metric (\ac{GSIM}) and Impedance Margin Ratio (\ac{IMR})  \cite{Henderson2024GridSystems,Zhu2024} &\vspace{-4mm}\begin{itemize}[leftmargin=*] \setlength\itemsep{-0.5em}
\item Can be used for both initial system screening to assess system strength as well as for further stability analysis from the information obtained 
\item The metrics are calculated at a range of frequency of interest hence, both converter control bandwidth is captured and hence applicable for OWPP clusters \end{itemize} &\vspace{-4mm}\begin{itemize}[leftmargin=*]\setlength\itemsep{-0.5em} 
\item The approach is only at the research level with no real project application   
\item More validation tests of the metrics are required \end{itemize} & Proposed metrics for obtaining system strength for a converter-dominated system \\
\hline
\end{tabular}
}
\label{tab:comparison_table1}
\end{table*}

%% Comparison of the methods
\begin{table*}[h!]
\tiny
\caption{Comparison of stability analysis methods}
\resizebox{\textwidth}{!}{%
\centering
\begin{tabular}{>{\raggedright\arraybackslash}p{3.0cm}>{\raggedright\arraybackslash}p{3.96cm}>{\raggedright\arraybackslash}p{3.96cm}>{\raggedright\arraybackslash}p{3.96cm}}
\hline
\textbf{Method} & \textbf{Advantages} & \textbf{Limitations} & \textbf{Practical Application} \\
\hline
Transfer function based  
\cite{Zhao2016VoltageGrid,Fan2019ModelingGrids,Amin2019AImpedance,Prieto-Araujo2011MethodologyFarms,Li2022ImpedanceApplication,Gong2018ParametricConverters}
&\vspace{-4mm}\begin{itemize}[leftmargin=*]\setlength\itemsep{-0.5em} 
\item It is helpful for converter control design, and the stability metric is well-defined. 
\item  Possible to build a numerical Transfer function via Vector Fitting. \end{itemize} &\vspace{-4mm}\begin{itemize}[leftmargin=*]\setlength\itemsep{-0.5em} 
\item Tedious to formulate complex transfer functions for large power systems. Thus, it is gradually being replaced by the impedance-based method. 
\item It is strictly related to an input/output relation. It is possible to derive it directly from state-space form. \end{itemize} & 
It has to be applied to the aggregated version of \acp{OWPP}. For accurate evaluations, the \ac{HVDC} or \ac{HVAC} cable multiple transfer functions of the $\pi$-section are required (in addition to filters and transformers). For point-to-point \ac{HVAC}, the method has to take into account also the cable. Conversely, for \ac{HVDC}-\ac{OWPP}, it is possible to avoid the cable since the \ac{AC} side is decoupled from the \ac{DC} one. 
\\
\hline
Impedance based \cite{Ji2020ImpedanceIntegration, Ghosh2019ImpedanceModel,Lyu2016FrequencyIntegration,Man2020Frequency-CouplingSystem,Zhou2021ASystems,Jacobs2023ElectricOnline,WGC4.492024Multi-frequencyBROCHURE} &\vspace{-4mm} 
\begin{itemize}[leftmargin=*]\setlength\itemsep{-0.5em} 
\item Powerful tool to evaluate stability at different phases of power system projects with well-defined metrics.
\item Black box models can be directly used. 
\item Offer reduced effort to screen multi-vendor interoperability\end{itemize}& \vspace{-4mm}\begin{itemize}[leftmargin=*]\setlength\itemsep{-0.5em}
    \item The impedance characteristic is dependent on the operating point for the lower frequency range.

    \item Trial \& Error approach on the parameters selection and tunning values for Phase Margin correction.
\end{itemize} & 
It can be used for any type of \acp{OWPP} dividing the study two-by-two. When the study is \ac{OWPP}-Grid, in a context of high \acp{IBR} penetration (low \ac{SCR}), might be preferable to use a frequency-domain representation rather than a Thevenin Equivalent.

\\

\hline
Passivity based  
\cite{Kocewiak2020OverviewSystems,HubertAalborgMMC-HVDC,Wu2023Passivity-BasedInverter,Wang2022PassivityDamping,Harnefors2017VSCAssessment,Harnefors2016Passivity-BasedOverview,Harnefors2008Frequency-domainDesign,WGC4.492024Multi-frequencyBROCHURE}
&\vspace{-4mm}\begin{itemize}[leftmargin=*]\setlength\itemsep{-0.5em} 
\item Useful for initial screening to extract information about non-passive regions of a converter. 
\item Guarantees stability on a wider frequency range than the impedance-based method. \end{itemize} &\vspace{-4mm}\begin{itemize}[leftmargin=*]\setlength\itemsep{-0.5em} 
\item Limited application within the low-frequency range 
\item Passivity does not imply that the system will guarantee good performance. \end{itemize} & 

The passivity method is becoming an interesting perspective from \acp{WT} and \ac{VSC}-\ac{MMC} manufacturers since once passivity is guaranteed in the specific frequency range, and virtually, stability is ensured. For the offshore side of a \ac{HVDC}-\ac{OWPP}, the passivity should be guaranteed just by converter manufacturers. On the onshore, it has to be guaranteed also by the \ac{TSO}.\\
\hline
Eigenvalue/Modal based   \cite{Amin2017Small-SignalMethods,Prieto-Araujo2011MethodologyFarms,KroutikovaState-SpaceMode,Yang2020UnifiedConverters,Ghosh2019ImpedanceModel,Kunjumuhammed2017TheSystems,WorkstreamWindEurope,Liao2023StabilitySystems,Misyris2022Zero-inertiaSharing,Stoorvogel2000TheApproach,Mesanovic2020ScienceDirectOptimization,WGC4.492024Multi-frequencyBROCHURE} &\vspace{-4mm}\begin{itemize}[leftmargin=*]\setlength\itemsep{-0.5em} 
\item The root cause of interaction/instability issues can be identified. 
\item Advanced controls and analysis are possible in this approach. \end{itemize} &\vspace{-4mm}\begin{itemize}[leftmargin=*]\setlength\itemsep{-0.5em} 
\item It is challenging to derive state-space models for converters, especially in the high-frequency range. 
\item Numerical or vector fitting generated state space does not have information about the states \end{itemize} & 
The method is useful in the preliminary design phase and tuning.  Where it is possible to identify how \ac{OWPP} parameters interact with \acp{WT}' \ac{VSC} controls and how it influences the stability. 
Identification and allocation of responsibility in multi-vendor context.
\\
\hline
\ac{EMT} Simulations  
\cite{ENTSO-E2021WorkstreamDevices,Watson2020AnTechnology,VordemBergeSOLUTIONSSYSTEMSb,O.D.ADEUYI2020Multi-terminalSchemes}
&\vspace{-4mm}\begin{itemize}[leftmargin=*]\setlength\itemsep{-0.5em} 
\item  The most accurate tool available to date and used to validate other methods.
\item Most suited for multi-frequency and large-signal stability analysis
\item System dynamics, including converter non-linearities, are well-represented. The response of a detailed dynamic model brings confidence in the results.
\item It is possible to use of \ac{IP} protected black boxed models directly from manufacturers  \end{itemize} &\vspace{-4mm}
\begin{itemize}[leftmargin=*]\setlength\itemsep{-0.5em}
    \item Time-consuming for accurate analysis
    \item It is not feasible for large systems. Therefore, study cases must be carefully selected.
    \item It cannot provide general information outside the studied cases.
    \item It can be challenging in some cases to identify the root cause of undesired response.
\end{itemize}
& Offline \ac{EMT} simulation has become a standard procedure for validating responses from small-signal stability analysis methods. It is also possible to perform a large-signal analysis. While it is common to practice \ac{EMT} analysis via black-box \acp{WT}' \ac{VSC} or \ac{MMC} converters, the grid side is still commonly a Thevenin generator. \\
\hline
Real-time Simulations 
\cite{P.Rault2018D9.3:Systems,RTDSTechnologies2022ENSURINGTESTING,VordemBergeSOLUTIONSSYSTEMSb,Mitra2014OffshoreStudy}
&\vspace{-4mm}\begin{itemize}[leftmargin=*]\setlength\itemsep{-0.5em} 
\item Useful for validation that takes into account hardware's effects. 
\item Provide a platform to test the behavior of the controller in a real environment before being deployed to the site. 
 \end{itemize} &\vspace{-4mm}\begin{itemize}[leftmargin=*]\setlength\itemsep{-0.5em} 
\item  Setting up the test center with C\&P replicas is expensive and might be difficult when more than two vendors are involved.
\item  Accuracy is proportional to the number of carried tests/contingencies \end{itemize} & The approach has already been applied in a real project (Johan Sverdrup) to analyze the control interaction of two \ac{HVDC} converters from different vendors.\\
\hline
Frequency sweep/scanning 
\cite{DasComparisonSystems,Jacobs2023AParks,Shirinzad2021FrequencySystems,Bakhshizadeh2016CouplingsConverters,Ren2016AFarms,Rygg2016ASystems,WGC4.492024Multi-frequencyBROCHURE,Pintelon2012SystemApproach,Liao2020ImpedanceConverters}
&\vspace{-4mm}\begin{itemize}[leftmargin=*]\setlength\itemsep{-0.5em} 
\item Frequency sweep is used to derive converter impedance for model validation, passivity evaluation, and impedance-based stability analysis. 
\item Numerical transfer function or state space models can also be obtained from the sweeps.\end{itemize} &\vspace{-4mm}\begin{itemize}[leftmargin=*]\setlength\itemsep{-0.5em} 
\item Requires detailed modeling and multiple sweep repetitions for multiple operational conditions, which is time-consuming. 
\item Shared curves by \acp{TSO} or manufacturers are valid just for specific operational points.
\end{itemize} & It is becoming the most preferred method for \ac{OWPP} stability analysis because of the possibility of obtaining impedance for black box models and for a wide frequency range. \\
\hline
\end{tabular}
}
\label{tab:comparison_table2}
\end{table*}

\section{Conclusion}
\label{sec:conclusion}
The interoperability of Offshore Wind Power Plant clusters and their interactions with a progressively weak grid has been recognized as a challenge by both the industry and academic community. Therefore, a detailed analysis has been conducted on different methods that could be used to investigate control interactions in converter-dominated power systems. Thereafter, a comparison of the methods based on their industry maturity, advantages, and limitations has been presented. Furthermore, the most appropriate methods to study control interactions have been discussed and evaluated depending on the type of \ac{OWPP}. Besides, their strengths and weaknesses have been highlighted.

Based on the analysis, it can be concluded that a single method is not sufficient to analyze all the stability issues. Therefore, a combination of both frequency domain and time domain simulation is needed. Regarding \acp{OWPP}, which can range from nearly to fully converter-based, the analysis should ensure that the converter dynamics, interactions among inverters, and other components are considered. Based on the type of \ac{OWPP}, it is advisable to adopt a suitable mix of stability analysis methods. In certain cases, different methods may be applied to the onshore and offshore sections of the \ac{OWPP}. The vision ahead is to establish more industry-standard practices for these methods.
%Hence, Time domain simulations employing Electromagnetic Transients (\ac{EMT}) frequency sweeps can be used to obtain linear models for frequency response analysis.

\section*{Acknowledgment}

This work is supported by the European Union as part of ADOreD project funded by the Horizon Europe MSCA programme (\href{https://www.msca-adored.eu/}{HORIZON-MSCA-2021-DN, Grant agreement 101073554})

\bibliographystyle{elsarticle-num}  
\bibliography{fullbib.bib}

\begin{thebibliography}{100}
\expandafter\ifx\csname url\endcsname\relax
  \def\url#1{\texttt{#1}}\fi
\expandafter\ifx\csname urlprefix\endcsname\relax\def\urlprefix{URL }\fi
\expandafter\ifx\csname href\endcsname\relax
  \def\href#1#2{#2} \def\path#1{#1}\fi

\bibitem{GlobalWindEnergyCouncil2024GLOBAL2024}
{Global Wind Energy Council}, \href{www.gwec.net}{Global offshore wind report 2024}, Tech. rep. (2024).
\newline\urlprefix\url{www.gwec.net}

\bibitem{PatentOffice2023OffshoreReport}
E.~Patent~Office, I.~Renewable Energy~Agency, {Offshore wind energy: Patent insight report}, Tech. rep. (2023).

\bibitem{RenewableEnergyAgency2023WorldPathway}
I.~Renewable Energy~Agency, \href{www.irena.org}{{World Energy Transitions Outlook 2023: 1.5{${}^\circ$}C Pathway}}, 2023.
\newline\urlprefix\url{www.irena.org}

\bibitem{Savaghebi2023GeneralPioneer}
L.~Savaghebi, M.~Zhang, W.~Keles, D.~Ladenburg, J.~Vest, M.~R. Seger, B.~Mortensen, N.~H. Sin, G.~Kitzing, L.~Kolios, .~. Uhd, {General rights Denmark as the Energy Island Pioneer}, Tech. rep. (2023).

\bibitem{Author2023ModelAnalysisb}
T.~Author, L.~Reis, J.~Are Wold~Suul, {Model Identification of Power Electronic Systems for Interaction Studies and Small-Signal Analysis}, Tech. rep. (2023).

\bibitem{RTEInteractionFarms}
{RTE}, \href{https://www.rte-international.com/en/references}{{Interaction Studies Between Sofia {\&} Dogger Bank C Offshore wind farms}} (2022).
\newline\urlprefix\url{https://www.rte-international.com/en/references}

\bibitem{AbdalrahmanAdil2016DolWin1Farms}
{Abdalrahman Adil}, {Isabegovic Emir}, {DolWin1: Challenges of Connecting Offshore Wind Farms}, IEEE International Energy Conference (ENERGYCON), Leuven, Belgium, 2016.

\bibitem{Lu2017LettersFilter}
M.~Lu, X.~Wang, P.~C. Loh, F.~Blaabjerg, \href{http://www.ieee.org/publications}{{Letters Resonance Interaction of Multiparallel Grid-Connected Inverters With LCL Filter}}, IEEE TRANSACTIONS ON POWER ELECTRONICS 32~(2) (2017).
\newblock \href {https://doi.org/10.1109/TPEL.2016.2585547} {\path{doi:10.1109/TPEL.2016.2585547}}.
\newline\urlprefix\url{http://www.ieee.org/publications}

\bibitem{TenneT2023AnnexesAgreement}
{TenneT}, \href{https://www.tennet.eu/}{{Annexes to Connection and Transmission Agreement}}, Tech. rep. (5 2023).
\newline\urlprefix\url{https://www.tennet.eu/}

\bibitem{WhatGuardian}
{What are the questions raised by the UK's recent blackout? | National Grid | The Guardian}.

\bibitem{Rudnik2022AnalysisSources}
V.~E. Rudnik, R.~A. Ufa, Y.~Y. Malkova, {Analysis of low-frequency oscillation in power system with renewable energy sources}, Energy Reports 8 (2022) 394--405.
\newblock \href {https://doi.org/10.1016/J.EGYR.2022.07.022} {\path{doi:10.1016/J.EGYR.2022.07.022}}.

\bibitem{BAKHSHIZADEH2019GridPlants}
M.~Bakhshizadeh, Å.~Kocewiak, J.~Hjerrild, F.~Laabjerg, C.~Leth~BAK, {Grid Converter Stability Aspects in Offshore Wind Power Plants}, in: CIGRE Symposium, CIGRE, Aalborg, 2019.

\bibitem{C.Buchhagen2016HarmonicTSO}
{C. Buchhagen}, {M. Greve}, {A. Menze}, {J. Jung}, {Harmonic Stability – Practical Experience of a TSO}, 15th International Workshop on Large-Scale Integration of Wind Power into Power Systems as well as Transmission Networks for Offshore Wind Farms, Vienna, 2016.

\bibitem{WGC4.492024Multi-frequencyBROCHURE}
{WG C4.49}, {Multi-frequency stability of converter-based modern power systems Power system technical performance C4 TECHNICAL BROCHURE}, Vol. 928, CIGRE, 2024.

\bibitem{RTDSTechnologies2022ENSURINGTESTING}
{RTDS Technologies}, \href{https://knowledge.rtds.com/hc/en-us/article_attachments/11123496721943}{{Ensuring interoperability for multi-vendor hvdc via replica testing}}, Tech. rep. (2022).
\newline\urlprefix\url{https://knowledge.rtds.com/hc/en-us/article_attachments/11123496721943}

\bibitem{VordemBergeSOLUTIONSSYSTEMS}
M.~Vor~dem Berge, S.~Denneti'ere, M.~Cai, \href{www.rte-international.com}{{Solutions and challenges to de-risk development of large-scale multi-vendor converter dominated systems}} (2022).
\newline\urlprefix\url{www.rte-international.com}

\bibitem{Elliott2016ABritain}
D.~Elliott, K.~R. W~Bell, S.~J. Finney, R.~Adapa, C.~Brozio, J.~Yu, K.~Hussain, \href{http://www.ieee.org/publications_standards/publications/rights/index.html}{{A Comparison of AC and HVDC Options for the Connection of Offshore Wind Generation in Great Britain}}, IEEE TRANSACTIONS ON POWER DELIVERY 31~(2) (2016).
\newblock \href {https://doi.org/10.1109/TPWRD.2015.2453233} {\path{doi:10.1109/TPWRD.2015.2453233}}.
\newline\urlprefix\url{http://www.ieee.org/publications_standards/publications/rights/index.html}

\bibitem{Dakic2021HVACTransmission}
J.~Dakic, M.~Cheah-Mane, O.~Gomis-Bellmunt, E.~Prieto-Araujo, {HVAC Transmission System for Offshore Wind Power Plants including Mid-Cable Reactive Power Compensation: Optimal Design and Comparison to VSC-HVDC Transmission}, IEEE Transactions on Power Delivery 36~(5) (2021) 2814--2824.
\newblock \href {https://doi.org/10.1109/TPWRD.2020.3027356} {\path{doi:10.1109/TPWRD.2020.3027356}}.

\bibitem{Gomis-Bellmunt2011Voltage-currentFarms}
O.~Gomis-Bellmunt, J.~Liang, J.~Ekanayake, N.~Jenkins, {Voltage-current characteristics of multiterminal HVDC-VSC for offshore wind farms}, Electric Power Systems Research 81~(2) (2011) 440--450.
\newblock \href {https://doi.org/10.1016/j.epsr.2010.10.007} {\path{doi:10.1016/j.epsr.2010.10.007}}.

\bibitem{Nazir2022Multi-terminalCoast}
M.~Nazir, J.~H. Enslin, E.~Hines, J.~D. McCalley, P.~A. Lof, B.~K. Garnick, {Multi-terminal HVDC Grid Topology for large Scale Integration of Offshore Wind on the U.S Atlantic Coast}, in: 2022 7th IEEE Workshop on the Electronic Grid, eGRID 2022, Institute of Electrical and Electronics Engineers Inc., 2022.
\newblock \href {https://doi.org/10.1109/eGRID57376.2022.9990011} {\path{doi:10.1109/eGRID57376.2022.9990011}}.

\bibitem{Liao2023StudyGrid}
Y.~Liao, K.~Xu, Y.~Varetsky, M.~Gajdzica, \href{https://doi.org/10.3390/en16020811}{{Study of Short Circuit and Inrush Current Impact on the Current-Limiting Reactor Operation in an Industrial Grid}} (2023).
\newblock \href {https://doi.org/10.3390/en16020811} {\path{doi:10.3390/en16020811}}.
\newline\urlprefix\url{https://doi.org/10.3390/en16020811}

\bibitem{Wiechowski2008SelectedTSO}
W.~Wiechowski, P.~B. Eriksen, {Selected studies on offshore wind farm cable connections - challenges and experience of the Danish TSO}, IEEE Power and Energy Society 2008 General Meeting: Conversion and Delivery of Electrical Energy in the 21st Century, PES (2008).
\newblock \href {https://doi.org/10.1109/PES.2008.4596124} {\path{doi:10.1109/PES.2008.4596124}}.

\bibitem{Wu2024GridTransmissionb}
D.~Wu, G.~S. Seo, L.~Xu, C.~Su, L.~Kocewiak, Y.~Sun, Z.~Qin, {Grid Integration of Offshore Wind Power: Standards, Control, Power Quality and Transmission}, IEEE Open Journal of Power Electronics 5 (2024) 583--604.
\newblock \href {https://doi.org/10.1109/OJPEL.2024.3390417} {\path{doi:10.1109/OJPEL.2024.3390417}}.

\bibitem{Nguyen2021ACondensers}
H.~T. Nguyen, M.~N. Chleirigh, G.~Yang, {A Technical Economic Evaluation of Inertial Response from Wind Generators and Synchronous Condensers}, IEEE Access 9 (2021) 7183--7192.
\newblock \href {https://doi.org/10.1109/ACCESS.2021.3049197} {\path{doi:10.1109/ACCESS.2021.3049197}}.

\bibitem{Rodrigues2015TrendsProjects}
S.~Rodrigues, C.~Restrepo, E.~Kontos, R.~Teixeira~Pinto, P.~Bauer, {Trends of offshore wind projects} (10 2015).
\newblock \href {https://doi.org/10.1016/j.rser.2015.04.092} {\path{doi:10.1016/j.rser.2015.04.092}}.

\bibitem{Ansari2020MMCSchemes}
J.~A. Ansari, C.~Liu, S.~A. Khan, {MMC based MTDC grids: A detailed review on issues and challenges for operation, control and protection schemes}, IEEE Access 8 (2020) 168154--168165.
\newblock \href {https://doi.org/10.1109/ACCESS.2020.3023544} {\path{doi:10.1109/ACCESS.2020.3023544}}.

\bibitem{Korompili2024ReviewIntegration}
A.~. Korompili, Q.~. Wu, H.~Zhao, \href{https://doi.org/10.1016/j.rser.2016.01.064}{{Review of VSC HVDC Connection for Offshore Wind Power Integration}}, Renewable {\&} Sustainable Energy Reviews 59 (2024) 1405--1414.
\newblock \href {https://doi.org/10.1016/j.rser.2016.01.064} {\path{doi:10.1016/j.rser.2016.01.064}}.
\newline\urlprefix\url{https://doi.org/10.1016/j.rser.2016.01.064}

\bibitem{Bathurst2015DeliverySystem}
G.~Bathurst, P.~Bordignan, {Delivery of the Nan'ao Multi-terminal VSC-HVDC System}, Tech. rep., IET (2 2015).

\bibitem{NRElectricCo.Ltd2015WorldsLinks}
{NR Electric Co. Ltd}, {World's First 5-Terminal VSC-HVDC Links}, Tech. rep. (2015).

\bibitem{CutululisGeneralSources}
N.~A.~. Cutululis, M.~. Koivisto, {Das}, \href{https://doi.org/10.11581/DTU.00000206}{{General rights Power systems of the future with very large shares of renewable energy sources}}\href {https://doi.org/10.11581/DTU.00000206} {\path{doi:10.11581/DTU.00000206}}.
\newline\urlprefix\url{https://doi.org/10.11581/DTU.00000206}

\bibitem{DCSolutions}
\href{www.twenties-project.eu}{{DC grids: motivation, feasibility and outstanding issues Status report for the European Commission Deliverable: D5.4 EC-GA n{\textordmasculine} 249812 Project full title: Transmission system operation with large penetration of Wind and other renewable Electricity sources in Networks by means of innovative Tools and Integrated Energy Solutions}}, Tech. rep.
\newline\urlprefix\url{www.twenties-project.eu}

\bibitem{ENTSO-E2021WorkstreamDevices}
{ENTSO-E}, {T{\&}D Europe}, {Workstream for the development of multi-vendor HVDC systems and other power electronics interfaced devices}, Tech. rep. (2021).

\bibitem{ENTSO-E2021EuropeanInteroperability}
{ENTSO-E}, \href{https://eepublicdownloads.entsoe.eu/clean-documents/Publications/Position%20papers%20and%20reports/210125_Offshore%20Development_Interoperability.pdf}{{European Network of Transmission System Operators for Electricity ENTSO-E Position on Offshore Development: Interoperability}}, Tech. rep. (2021).
\newline\urlprefix\url{https://eepublicdownloads.entsoe.eu/clean-documents/Publications/Position%20papers%20and%20reports/210125_Offshore%20Development_Interoperability.pdf}

\bibitem{GridAmendment}
{Grid Connection European Stakeholder Committee Expert Group on Connection Requirements for Offshore Systems-Phase II (Proposal for the NC HVDC amendment)}.

\bibitem{EuropeanClimate2023SuccessfulProject}
I.~European~Climate, E.~E. Agency, {Successful kick-off of InterOPERA Horizon Europe offshore electricity grids project}, Tech. rep. (1 2023).

\bibitem{Conseilinternationaldesgrandsreseauxelectriques.ComitedetudesB4.2008SystemsInfeed}
{Conseil international des grands r'eseaux 'electriques. Comit'e d''etudes B4.}, {Impr. Conformes)}, {Systems with multiple DC infeed}, CIGR'e, 2008.

\bibitem{Henderson2024GridSystems}
C.~Henderson, A.~Egea-Alvarez, T.~Kneuppel, G.~Yang, L.~Xu, {Grid Strength Impedance Metric: An Alternative to SCR for Evaluating System Strength in Converter Dominated Systems}, IEEE Transactions on Power Delivery 39~(1) (2024) 386--396.
\newblock \href {https://doi.org/10.1109/TPWRD.2022.3233455} {\path{doi:10.1109/TPWRD.2022.3233455}}.

\bibitem{AEMC2017}
A.~E. M.~C. (AEMC), \href{https://www.aemc.gov.au/sites/default/files/content/4645acea-e66f-4b5b-94a1-1dd14e7f8a93/ERC0211-Final-determination.pdf}{National electricity amendment (managing power system fault levels) rule 2017} (2017).
\newline\urlprefix\url{https://www.aemc.gov.au/sites/default/files/content/4645acea-e66f-4b5b-94a1-1dd14e7f8a93/ERC0211-Final-determination.pdf}

\bibitem{system_ISGT}
A.~Bori\v{c}i\'c, J.~L.~R. Torres, M.~Popov, System strength: Classification, evaluation methods, and emerging challenges in ibr-dominated grids, in: Proceedings of the 11th International Conference on Innovative Smart Grid Technologies - Asia, ISGT-Asia 2022, Institute of Electrical and Electronics Engineers Inc., 2022, pp. 185--189.
\newblock \href {https://doi.org/10.1109/ISGTAsia54193.2022.10003499} {\path{doi:10.1109/ISGTAsia54193.2022.10003499}}.

\bibitem{Henderson2022AnalysisSystems}
C.~Henderson, A.~Egea-Alvarez, L.~Xu, {Analysis of multi-converter network impedance using MIMO stability criterion for multi-loop systems}, Electric Power Systems Research 211 (10 2022).
\newblock \href {https://doi.org/10.1016/j.epsr.2022.108542} {\path{doi:10.1016/j.epsr.2022.108542}}.

\bibitem{Zhu2024}
Y.~Zhu, T.~C. Green, X.~Zhou, Y.~Li, D.~Kong, Y.~Gu, Impedance margin ratio: a new metric for small-signal system strength, IEEE Transactions on Power Systems (2024).
\newblock \href {https://doi.org/10.1109/TPWRS.2024.3371231} {\path{doi:10.1109/TPWRS.2024.3371231}}.

\bibitem{Ghimire2023Small-SignalCondensers}
S.~Ghimire, K.~V. Kkuni, E.~D. Guest, K.~H. Jensen, G.~Yang, \href{http://arxiv.org/abs/2310.06457}{{Small-Signal Stability and SCR Enhancement of Offshore WPPs with Synchronous Condensers}} (10 2023).
\newline\urlprefix\url{http://arxiv.org/abs/2310.06457}

\bibitem{Henderson2022ExploringConverters}
C.~Henderson, A.~Egea-Alvarez, P.~Papadopoulos, R.~Li, L.~Xu, R.~Da~Silva, A.~Kinsella, I.~Gutierrez, R.~Pabat-Stroe, {Exploring an Impedance-Based SCR for Accurate Representation of Grid-Forming Converters}, in: IEEE Power and Energy Society General Meeting, Vol. 2022-July, IEEE Computer Society, 2022.
\newblock \href {https://doi.org/10.1109/PESGM48719.2022.9916733} {\path{doi:10.1109/PESGM48719.2022.9916733}}.

\bibitem{Conseilinternationaldesgrandsreseauxelectriques.ComitedetudesB4.2016ConnectionNetworks}
{Conseil international des grands r'eseaux 'electriques. Comit'e d''etudes B4.}, {Impr. Conformes)}, {Connection of wind farms to weak AC networks}, CIGR'e, 2016.

\bibitem{NERC2017NERCSystems}
{NERC}, {NERC | Report Title | Report Date I Integrating Inverter-Based Resources into Low Short Circuit Strength Systems}, Tech. rep. (12 2017).

\bibitem{Evaluate2014}
Y.~Zhang, S.-H.~F. Huang, J.~Schmall, J.~Conto, J.~Billo, E.~Rehman, Evaluating system strength for large-scale wind plant integration, in: 2014 IEEE PES General Meeting | Conference \& Exposition, 2014, pp. 1--5.
\newblock \href {https://doi.org/10.1109/PESGM.2014.6939043} {\path{doi:10.1109/PESGM.2014.6939043}}.

\bibitem{Davari2017RobustDynamics}
M.~Davari, Y.~A. R.~I. Mohamed, {Robust Vector Control of a Very Weak-Grid-Connected Voltage-Source Converter Considering the Phase-Locked Loop Dynamics}, IEEE Transactions on Power Electronics 32~(2) (2017) 977--994.
\newblock \href {https://doi.org/10.1109/TPEL.2016.2546341} {\path{doi:10.1109/TPEL.2016.2546341}}.

\bibitem{Harnefors2007ModelingMatrices}
L.~Harnefors, {Modeling of three-phase dynamic systems using complex transfer functions and transfer matrices}, IEEE Transactions on Industrial Electronics 54~(4) (2007) 2239--2248.
\newblock \href {https://doi.org/10.1109/TIE.2007.894769} {\path{doi:10.1109/TIE.2007.894769}}.

\bibitem{Zhang2019StabilityModeling}
H.~Zhang, L.~Harnefors, X.~Wang, H.~Gong, J.~P. Hasler, {Stability Analysis of Grid-Connected Voltage-Source Converters Using SISO Modeling}, IEEE Transactions on Power Electronics 34~(8) (2019) 8104--8117.
\newblock \href {https://doi.org/10.1109/TPEL.2018.2878930} {\path{doi:10.1109/TPEL.2018.2878930}}.

\bibitem{Bakhshizadeh2016CouplingsConverters}
M.~K. Bakhshizadeh, X.~Wang, F.~Blaabjerg, J.~Hjerrild, L.~Kocewiak, C.~L. Bak, B.~Hesselbak, {Couplings in Phase Domain Impedance Modeling of Grid-Connected Converters}, IEEE Transactions on Power Electronics 31~(10) (2016) 6792--6796.
\newblock \href {https://doi.org/10.1109/TPEL.2016.2542244} {\path{doi:10.1109/TPEL.2016.2542244}}.

\bibitem{Zhao2016VoltageGrid}
M.~Zhao, X.~Yuan, J.~Hu, Y.~Yan, {Voltage Dynamics of Current Control Time-Scale in a VSC-Connected Weak Grid}, IEEE Transactions on Power Systems 31~(4) (2016) 2925--2937.
\newblock \href {https://doi.org/10.1109/TPWRS.2015.2482605} {\path{doi:10.1109/TPWRS.2015.2482605}}.

\bibitem{Fan2019ModelingGrids}
L.~Fan, {Modeling Type-4 Wind in Weak Grids}, IEEE Transactions on Sustainable Energy 10~(2) (2019) 853--864.
\newblock \href {https://doi.org/10.1109/TSTE.2018.2849849} {\path{doi:10.1109/TSTE.2018.2849849}}.

\bibitem{Amin2019AImpedance}
M.~Amin, M.~Molinas, {A gray-box method for stability and controller parameter estimation in hvdc-connected wind farms based on nonparametric impedance}, IEEE Transactions on Industrial Electronics 66~(3) (2019) 1872--1882.
\newblock \href {https://doi.org/10.1109/TIE.2018.2840516} {\path{doi:10.1109/TIE.2018.2840516}}.

\bibitem{Prieto-Araujo2011MethodologyFarms}
E.~Prieto-Araujo, F.~D. Bianchi, A.~Junyent-Ferr{\'{e}}, O.~Gomis-Bellmunt, {Methodology for droop control dynamic analysis of multiterminal VSC-HVDC grids for offshore wind farms}, IEEE Transactions on Power Delivery 26~(4) (2011) 2476--2485.
\newblock \href {https://doi.org/10.1109/TPWRD.2011.2144625} {\path{doi:10.1109/TPWRD.2011.2144625}}.

\bibitem{Li2022ImpedanceApplication}
Y.~Li, Q.~Liu, B.~Wu, Y.~Wang, Q.~Huang, {Impedance Transfer Functions Fitting Methods of Grid-connected Inverters: Comparison and Application}, 2022 International Conference on Power Energy Systems and Applications, ICoPESA 2022 (2022) 331--336\href {https://doi.org/10.1109/ICOPESA54515.2022.9754427} {\path{doi:10.1109/ICOPESA54515.2022.9754427}}.

\bibitem{DArco2019Time-Invariant-sections}
S.~D'Arco, J.~A. Suul, J.~Beerten, {Time-Invariant State-Space model of an AC Cable by dq-representation of Frequency-Dependent {$\pi$}-sections}, Proceedings of 2019 IEEE PES Innovative Smart Grid Technologies Europe, ISGT-Europe 2019 (9 2019).
\newblock \href {https://doi.org/10.1109/ISGTEUROPE.2019.8905577} {\path{doi:10.1109/ISGTEUROPE.2019.8905577}}.

\bibitem{Gong2018ParametricConverters}
H.~Gong, D.~Yang, X.~Wang, {Parametric Identification of DQ Impedance Model for Three-Phase Voltage-Source Converters}, Proceedings - 2018 IEEE International Power Electronics and Application Conference and Exposition, PEAC 2018 (12 2018).
\newblock \href {https://doi.org/10.1109/PEAC.2018.8590653} {\path{doi:10.1109/PEAC.2018.8590653}}.

\bibitem{J.M.UndrillT.E.Kostyniak1976SubsynchronousAnalysis}
{J.M. Undrill; T.E. Kostyniak}, {Subsynchronous oscillations, part 1-Comprehensive system stability analysis}, Tech. rep., IEEE Trans. Power App. Syst., vol. PAS–95 (8 1976).

\bibitem{Liao2020Impedance-BasedPoles}
Y.~Liao, X.~Wang, {Impedance-Based Stability Analysis for Interconnected Converter Systems with Open-Loop RHP Poles}, IEEE Transactions on Power Electronics 35~(4) (2020) 4388--4397.
\newblock \href {https://doi.org/10.1109/TPEL.2019.2939636} {\path{doi:10.1109/TPEL.2019.2939636}}.

\bibitem{Wang2017FrequencySystem}
Y.~Wang, X.~Wang, F.~Blaabjerg, Z.~Chen, {Frequency scanning-based stability analysis method for grid-connected inverter system}, 2017 IEEE 3rd International Future Energy Electronics Conference and ECCE Asia, IFEEC - ECCE Asia 2017 (2017) 1575--1580\href {https://doi.org/10.1109/IFEEC.2017.7992281} {\path{doi:10.1109/IFEEC.2017.7992281}}.

\bibitem{Ghosh2019ImpedanceModel}
S.~Ghosh, K.~V. Kkuni, G.~Yang, L.~Kocewiak, {Impedance scan and characterization of Type 4 wind power plants through aggregated model}, IECON Proceedings (Industrial Electronics Conference) 2019-October (2019) 1799--1804.
\newblock \href {https://doi.org/10.1109/IECON.2019.8926629} {\path{doi:10.1109/IECON.2019.8926629}}.

\bibitem{Lyu2016FrequencyIntegration}
J.~Lyu, X.~Cai, M.~Molinas, {Frequency Domain Stability Analysis of MMC-Based HVdc for Wind Farm Integration}, IEEE Journal of Emerging and Selected Topics in Power Electronics 4~(1) (2016) 141--151.
\newblock \href {https://doi.org/10.1109/JESTPE.2015.2498182} {\path{doi:10.1109/JESTPE.2015.2498182}}.

\bibitem{Man2020Frequency-CouplingSystem}
J.~Man, X.~Xie, S.~Xu, C.~Zou, C.~Yin, {Frequency-Coupling Impedance Model Based Analysis of a High-Frequency Resonance Incident in an Actual MMC-HVDC System}, IEEE Transactions on Power Delivery 35~(6) (2020) 2963--2971.
\newblock \href {https://doi.org/10.1109/TPWRD.2020.3022504} {\path{doi:10.1109/TPWRD.2020.3022504}}.

\bibitem{Zhou2021ASystems}
W.~Zhou, R.~E. Torres-Olguin, Y.~Wang, Z.~Chen, {A Gray-Box Hierarchical Oscillatory Instability Source Identification Method of Multiple-Inverter-Fed Power Systems}, IEEE Journal of Emerging and Selected Topics in Power Electronics 9~(3) (2021) 3095--3113.
\newblock \href {https://doi.org/10.1109/JESTPE.2020.2992225} {\path{doi:10.1109/JESTPE.2020.2992225}}.

\bibitem{Jacobs2023ElectricOnline}
K.~Jacobs, Y.~Seyedi, L.~Meng, U.~Karaagac, J.~Mahseredjian, \href{https://doi.org/10.1016/j.epsr.2023.109311}{{Electric Power Systems Research 220 (2023) 109311 Available online}} (2023) 378--7796\href {https://doi.org/10.1016/j.epsr.2023.109311} {\path{doi:10.1016/j.epsr.2023.109311}}.
\newline\urlprefix\url{https://doi.org/10.1016/j.epsr.2023.109311}

\bibitem{Liao2018GeneralAnalysis}
Y.~Liao, X.~Wang, {General Rules of Using Bode Plots for Impedance-Based Stability Analysis}, 2018 IEEE 19th Workshop on Control and Modeling for Power Electronics, COMPEL 2018 (9 2018).
\newblock \href {https://doi.org/10.1109/COMPEL.2018.8460168} {\path{doi:10.1109/COMPEL.2018.8460168}}.

\bibitem{Dhua2017HarmonicPlants}
D.~. Dhua, G.~. Yang, Z.~. Zhang, Å.~H. Kocewiak, A.~Timofejevs, {Harmonic Active Filtering and Impedance-based Stability Analysis in Offshore Wind Power Plants}, Citation (2017).

\bibitem{Liao2020ImpedanceConverters}
Y.~Liao, {Impedance Modeling and Stability Analysis of Grid-Interactive Converters} (2020).
\newblock \href {https://doi.org/10.5278/VBN.PHD.ENG.00086} {\path{doi:10.5278/VBN.PHD.ENG.00086}}.

\bibitem{Khalil2014NonlinearSystems}
H.~K.~. Khalil, {Nonlinear systems} (2014).

\bibitem{Wu2023Passivity-BasedInverter}
G.~Wu, Y.~He, H.~Zhang, X.~Wang, D.~Pan, X.~Ruan, C.~Yao, {Passivity-Based Stability Analysis and Generic Controller Design for Grid-Forming Inverter}, IEEE Transactions on Power Electronics 38~(5) (2023) 5832--5843.
\newblock \href {https://doi.org/10.1109/TPEL.2023.3237608} {\path{doi:10.1109/TPEL.2023.3237608}}.

\bibitem{Kocewiak2020OverviewSystems}
Å.~Kocewiak, R.~Blasco-Gimenez, C.~Buchhagen, {Overview, Status and Outline of Stability Analysis in Converter-based Power Systems}, Tech. rep., 19th Int'l Wind Integration Workshop (11 2020).

\bibitem{HubertAalborgMMC-HVDC}
Å.~Hubert, H.~Wu, X.~Wang, Å.~HKocewiak, J.~Hjerrild, M.~Kazem Bakhshizadeh Wind~Power, Ã.~Fredericia, {Aalborg Universitet Passivity-Based Harmonic Stability Analysis of an Offshore Wind Farm Connected to a MMC-HVDC}.

\bibitem{Wang2022PassivityDamping}
X.~Wang, Y.~He, D.~Pan, H.~Zhang, Y.~Ma, X.~Ruan, {Passivity Enhancement for LCL-Filtered Inverter With Grid Current Control and Capacitor Current Active Damping}, IEEE Transactions on Power Electronics 37~(4) (2022) 3801--3812.
\newblock \href {https://doi.org/10.1109/TPEL.2021.3111677} {\path{doi:10.1109/TPEL.2021.3111677}}.

\bibitem{Akhavan2021PassivityVariations}
A.~Akhavan, S.~Golestan, J.~C. Vasquez, J.~M. Guerrero, {Passivity Enhancement of Voltage-Controlled Inverters in Grid-Connected Microgrids Considering Negative Aspects of Control Delay and Grid Impedance Variations}, IEEE Journal of Emerging and Selected Topics in Power Electronics 9~(6) (2021) 6637--6649.
\newblock \href {https://doi.org/10.1109/JESTPE.2021.3065671} {\path{doi:10.1109/JESTPE.2021.3065671}}.

\bibitem{Liao2020Passivity-basedConverters}
Y.~Liao, X.~Wang, F.~Blaabjerg, {Passivity-based analysis and design of linear voltage controllers for voltage-source converters}, IEEE Open Journal of the Industrial Electronics Society 1~(1) (2020) 114--126.
\newblock \href {https://doi.org/10.1109/OJIES.2020.3001406} {\path{doi:10.1109/OJIES.2020.3001406}}.

\bibitem{Xie2020Passivity-BasedModel}
C.~Xie, K.~Li, J.~Zou, J.~M. Guerrero, {Passivity-Based Stabilization of LCL-Type Grid-Connected Inverters via a General Admittance Model}, IEEE Transactions on Power Electronics 35~(6) (2020) 6636--6648.
\newblock \href {https://doi.org/10.1109/TPEL.2019.2955861} {\path{doi:10.1109/TPEL.2019.2955861}}.

\bibitem{Wu2020Virtual-Flux-BasedVSCs}
H.~Wu, X.~Wang, {Virtual-Flux-Based Passivation of Current Control for Grid-Connected VSCs}, IEEE Transactions on Power Electronics 35~(12) (2020) 12673--12677.
\newblock \href {https://doi.org/10.1109/TPEL.2020.2997876} {\path{doi:10.1109/TPEL.2020.2997876}}.

\bibitem{Harnefors2017VSCAssessment}
L.~Harnefors, R.~Finger, X.~Wang, H.~Bai, F.~Blaabjerg, {VSC Input-Admittance Modeling and Analysis above the Nyquist Frequency for Passivity-Based Stability Assessment}, IEEE Transactions on Industrial Electronics 64~(8) (2017) 6362--6370.
\newblock \href {https://doi.org/10.1109/TIE.2017.2677353} {\path{doi:10.1109/TIE.2017.2677353}}.

\bibitem{Harnefors2016Passivity-BasedOverview}
L.~Harnefors, X.~Wang, A.~G. Yepes, F.~Blaabjerg, {Passivity-Based Stability Assessment of Grid-Connected VSCs-An Overview}, IEEE Journal of Emerging and Selected Topics in Power Electronics 4~(1) (2016) 116--125.
\newblock \href {https://doi.org/10.1109/JESTPE.2015.2490549} {\path{doi:10.1109/JESTPE.2015.2490549}}.

\bibitem{Harnefors2008Frequency-domainDesign}
L.~Harnefors, L.~Zhang, M.~Bongiorno, {Frequency-domain passivity-based current controller design}, IET Power Electronics 1~(4) (2008) 455--465.
\newblock \href {https://doi.org/10.1049/iet-pel:20070286} {\path{doi:10.1049/iet-pel:20070286}}.

\bibitem{Dorf2017ModernEdition}
R.~C. Dorf, R.~H. Bishop, {Modern Control Systems - Global Edition} (2017).

\bibitem{P.Kundur1994PowerControl}
{P. Kundur}, {N. J. Balu}, {M. G. Lauby}, {Power System Stability and Control}, McGraw-hill, New York, 1994.

\bibitem{M.J.Gibbard2015Small-signalSystems}
{M.J. Gibbard}, {P. Pourbeik}, {D.J. Vowles}, {Small-signal stability, control and dynamic performance of power systems}, University of Adelaide Press, Adelaide, 2015.

\bibitem{MatsLarsson2021SystematicAssessment}
{Mats Larsson}, {Systematic Approach to Harmonic Stability Assessment}, Tech. rep., Hitachi ABB Power Grids R{\_}D (2021).

\bibitem{Kocewiak2022PracticalMitigation}
L.~Kocewiak, R.~Blasco-Gimenez, C.~Buchhagen, J.~B. Kwon, M.~Larsson, Y.~Sun, X.~Wang, {Practical Aspects of Small-signal Stability Analysis and Instability Mitigation}, IET Conference Proceedings 2022~(23) (2022) 138--150.
\newblock \href {https://doi.org/10.1049/ICP.2022.2748} {\path{doi:10.1049/ICP.2022.2748}}.

\bibitem{Amin2017Small-SignalMethods}
M.~Amin, M.~Molinas, {Small-Signal Stability Assessment of Power Electronics Based Power Systems: A Discussion of Impedance-and Eigenvalue-Based Methods}, IEEE Transactions on Industry Applications 53~(5) (2017) 5014--5030.
\newblock \href {https://doi.org/10.1109/TIA.2017.2712692} {\path{doi:10.1109/TIA.2017.2712692}}.

\bibitem{KroutikovaState-SpaceMode}
N.~Kroutikova, C.~A. Hernandez-Aramburo, T.~C. Green, {State-Space Model of Grid-Connected Inverters under Current Control Mode}.

\bibitem{Yang2020UnifiedConverters}
D.~Yang, X.~Wang, {Unified Modular State-Space Modeling of Grid-Connected Voltage-Source Converters}, IEEE Transactions on Power Electronics 35~(9) (2020) 9702--9717.
\newblock \href {https://doi.org/10.1109/TPEL.2020.2965941} {\path{doi:10.1109/TPEL.2020.2965941}}.

\bibitem{Kunjumuhammed2017TheSystems}
L.~P. Kunjumuhammed, B.~C. Pal, C.~Oates, K.~J. Dyke, {The Adequacy of the Present Practice in Dynamic Aggregated Modeling of Wind Farm Systems}, IEEE Transactions on Sustainable Energy 8~(1) (2017) 23--32.
\newblock \href {https://doi.org/10.1109/TSTE.2016.2563162} {\path{doi:10.1109/TSTE.2016.2563162}}.

\bibitem{Liao2023StabilitySystems}
Y.~Liao, H.~Wu, X.~Wang, M.~Ndreko, R.~Dimitrovski, W.~Winter, {Stability and Sensitivity Analysis of Multi-Vendor, Multi-Terminal HVDC Systems}, IEEE Open Journal of Power Electronics 4 (2023) 52--66.
\newblock \href {https://doi.org/10.1109/OJPEL.2023.3234803} {\path{doi:10.1109/OJPEL.2023.3234803}}.

\bibitem{Conseilinternationaldesgrandsreseauxelectriques.JointworkinggroupC4-C6352018ModellingStudies}
{Conseil international des grands r'eseaux 'electriques. Joint working group C4-C6 (35)}, {Congr'es international des r'eseaux 'electriques de distribution.}, {Impr. Chauveau)}, {Modelling of inverter-based generation for power system dynamic studies}, CIGR'e, 2018.

\bibitem{Watson2020AnTechnology}
N.~R. Watson, J.~D. Watson, {An Overview of HVDC Technology}, Energies 2020, Vol. 13, Page 4342 13~(17) (2020) 4342.
\newblock \href {https://doi.org/10.3390/EN13174342} {\path{doi:10.3390/EN13174342}}.

\bibitem{P.Rault2018D9.3:Systems}
{P. Rault}, {O. Despouys}, {D9.3: BEST PATHS DEMO{\#}2 Final Recommendations For Interoperability Of Multivendor HVDC Systems}, Tech. rep. (12 2018).

\bibitem{Mitra2014OffshoreStudy}
P.~Mitra, L.~Zhang, L.~Harnefors, {Offshore wind integration to a weak grid by VSC-HVDC links using power-synchronization control: A case study}, IEEE Transactions on Power Delivery 29~(1) (2014) 453--461.
\newblock \href {https://doi.org/10.1109/TPWRD.2013.2273979} {\path{doi:10.1109/TPWRD.2013.2273979}}.

\bibitem{O.D.ADEUYI2020Multi-terminalSchemes}
{O.D. Adeuyi}, {M. H. Rahman}, {I. Cowan}, {B. Ponnalagan}, {Multi-terminal Extension of Embedded Point-to-Point VSC-HVDC Schemes}, Tech. rep., CIGRE B4-120 (9 2020).

\bibitem{Shirinzad2021FrequencySystems}
M.~Shirinzad, {Frequency Scan Based Stability Analysis of Power Electronic Systems}, Tech. rep. (2021).

\bibitem{Rik2012SystemApproach}
P.~Rik, S.~Johan, \href{https://ieeexplore.ieee.org/book/6198969}{{System Identification: A Frequency Domain Approach}}, 2nd Edition, Wiley-IEEE Press, 2012.
\newline\urlprefix\url{https://ieeexplore.ieee.org/book/6198969}

\bibitem{Rygg2018Impedance-basedSystems}
A.~Rygg, {Impedance-based Methods for Small-signal Analysis of Power Electronics Dominated Systems}, Ph.D. thesis, Norwegian University of Science and Technology (2018).

\bibitem{Rygg2016ASystems}
A.~Rygg, M.~Molinas, C.~Zhang, X.~Cai, {A Modified Sequence-Domain Impedance Definition and Its Equivalence to the dq-Domain Impedance Definition for the Stability Analysis of AC Power Electronic Systems}, IEEE Journal of Emerging and Selected Topics in Power Electronics 4~(4) (2016) 1383--1396.
\newblock \href {https://doi.org/10.1109/JESTPE.2016.2588733} {\path{doi:10.1109/JESTPE.2016.2588733}}.

\bibitem{DasComparisonSystems}
M.~K. Das, A.~M. Kulkarni, P.~B. Darji, {Comparison of DQ and Dynamic Phasor based Frequency Scanning Analysis of Grid-connected Power Electronic Systems}, Tech. rep.

\bibitem{Jiang1995ASYSTEMS}
X.~Jiang, A.~M. Gole, {A frequency scanning method for the identification of harmonic instabilities in hvdc systems}, Tech. Rep.~4 (1995).

\bibitem{Nouri2021TestPerturbations}
B.~Nouri, Å.~Kocewiak, S.~Shah, P.~Koralewicz, V.~Gevorgian, P.~S{\o}rensen, {Test methodology for validation of multi-frequency models of renewable energy generators using small-signal perturbations}, IET Renewable Power Generation 15~(15) (2021) 3564--3576.
\newblock \href {https://doi.org/10.1049/rpg2.12245} {\path{doi:10.1049/rpg2.12245}}.

\bibitem{Pintelon2012SystemApproach}
R.~R. Pintelon, J.~J. Schoukens, \href{https://researchportal.vub.be/en/publications/system-identification-a-frequency-domain-approach-2}{{System Identification: A frequency Domain Approach}}, Wiley / IEEE Press, 2012.
\newline\urlprefix\url{https://researchportal.vub.be/en/publications/system-identification-a-frequency-domain-approach-2}

\bibitem{KimpianMultiphasePractice}
T.~Kimpi{\'{a}}n, F.~Augusztinovicz, {Multiphase multisine signals-Theory and practice}.

\bibitem{Khan2024EnhancedSequence}
M.~A. Khan, F.~Taghizadeh, J.~Lu, F.~Bai, {Enhanced Online Impedance Estimation of Grid-Connected Inverter Using Hybrid Pseudorandom Binary Sequence}, 2024 International Conference on Green Energy, Computing and Sustainable Technology, GECOST 2024 (2024) 110--115\href {https://doi.org/10.1109/GECOST60902.2024.10475079} {\path{doi:10.1109/GECOST60902.2024.10475079}}.

\bibitem{Jacobs2023AParks}
K.~Jacobs, Y.~Seyedi, L.~Meng, U.~Karaagac, J.~Mahseredjian, {A comparative study on frequency scanning techniques for stability assessment in power systems incorporating wind parks}, Electric Power Systems Research 220 (2023) 109311.
\newblock \href {https://doi.org/10.1016/J.EPSR.2023.109311} {\path{doi:10.1016/J.EPSR.2023.109311}}.

\bibitem{Ren2016AFarms}
W.~Ren, E.~Larsen, {A Refined Frequency Scan Approach to Sub-Synchronous Control Interaction (SSCI) Study of Wind Farms}, IEEE Transactions on Power Systems 31~(5) (2016) 3904--3912.
\newblock \href {https://doi.org/10.1109/TPWRS.2015.2501543} {\path{doi:10.1109/TPWRS.2015.2501543}}.

\bibitem{PulseXplore}
\href{https://ieeexplore.ieee.org/book/5264450}{{Pulse Width Modulation for Power Converters: Principles and Practice | IEEE eBooks | IEEE Xplore}}.
\newline\urlprefix\url{https://ieeexplore.ieee.org/book/5264450}

\bibitem{Ji2020ImpedanceIntegration}
K.~Ji, G.~Tang, H.~Pang, J.~Yang, {Impedance Modeling and Analysis of MMC-HVDC for Offshore Wind Farm Integration}, IEEE Transactions on Power Delivery 35~(3) (2020) 1488--1501.
\newblock \href {https://doi.org/10.1109/TPWRD.2019.2946450} {\path{doi:10.1109/TPWRD.2019.2946450}}.

\bibitem{WorkstreamWindEurope}
{Workstream for the development of multi-vendor HVDC systems | WindEurope}.

\bibitem{Misyris2022Zero-inertiaSharing}
G.~S. Misyris, A.~Tosatto, S.~Chatzivasileiadis, T.~Weckesser, \href{https://doi.org/10.1109/TPWRS.2021.3113274}{{Zero-inertia Offshore Grids: N-1 Security and Active Power Sharing}}, IEEE Transactions on Power Systems 37~(3) (2022) 2052--2062.
\newblock \href {https://doi.org/10.1109/TPWRS.2021.3113274} {\path{doi:10.1109/TPWRS.2021.3113274}}.
\newline\urlprefix\url{https://doi.org/10.1109/TPWRS.2021.3113274}

\bibitem{Stoorvogel2000TheApproach}
A.~A. Stoorvogel, {The H $\infty$ control problem: a state space approach} (2000).

\bibitem{Mesanovic2020ScienceDirectOptimization}
A.~M. Me{\v{s}}anovi'c, U.~M{\"{u}}nz, R.~Findeisen, \href{www.sciencedirect.com}{{ScienceDirect ScienceDirect Controller tuning in power systems using singular value optimization}}, IFAC PapersOnLine 53~(2) (2020) 13501--13507.
\newblock \href {https://doi.org/10.1016/j.ifacol.2020.12.755} {\path{doi:10.1016/j.ifacol.2020.12.755}}.
\newline\urlprefix\url{www.sciencedirect.com}

\end{thebibliography}

\end{document}